\documentclass[twocolumn]{aa} 
\usepackage{graphicx}
\usepackage{txfonts}
\begin{document}
   \title{Structure and evolution of rotationally and tidally distorted stars}

   \subtitle{}

   \author{H.F. Song
          \inst{1,2}\fnmsep\thanks{Send offprint requests to: H. F. Song, e-mail: sci.hfsong@gzu.edu.cn}
          Z. Zhong\inst{1}
           \and Y.Lu\inst{1}
          }

   \institute{College of science, Guizhou University,
             Guiyang, 550025, P.R. China\\
              \email{sci.hfsong@gzu.edu.cn, zhongzhen307@yahoo.cn}
         \and
        Joint Centre for Astronomy, National Astronomical
        Observatories-Guizhou University ,
             Guiyang, 550025, P.R. China\\
            \email{songhanfeng@163.com}
             }

   \date{}


  \abstract
   {}
   {This paper aims to study the configuration of two components caused by
    rotational and tidal distortions in the model of a binary system.}
{The potentials of the two distorted components can be approximated
to 2nd-degree harmonics. Furthermore, both the accretion luminosity
($\sigma_{i}$) and the irradiative luminosity are included in
stellar structure equations.}
{The equilibrium structure of rotationally and tidally distorted
star is exactly a triaxial ellipsoids. A formula describing the
isobars is presented, and the rotational velocity and the
gravitational acceleration at the primary surface simulated. The
results show the distortion at the outer layers of the primary
increases with temporal variation and system evolution. Besides, it
was observed that the luminosity accretion is unstable, and the
curve of the energy-generation rate fluctuates after the main
sequence in rotation sequences. The luminosity in rotation sequences
is slightly weaker than that in non-rotation sequences. As a result,
the volume expands slowly. Polar ejection is intensified by the
tidal effect. The ejection of an equatorial ring may be favoured by
both the opacity effect and the $g_{e}(\theta,\varphi)$-effect in
the binary system.}
   {}

   \maketitle
%

\section{Introduction}
In the conventional model of binary stars, there is no consideration
of spin and tidal effects (Eggleton 1971,1972,1973; Hofmeister,
Kippenhahn \& Weigert, 1964; Kippenhahn et al. 1967; etc.); however,
rotation and tide have been regarded as two important physical
factors in recent years, so they need to be considered for a better
understanding of the evolution of massive close binaries (e.g.,
Heger, Langer \& Woosley 2000a; Meynet \& Maeder 2000). The
structure and evolution of rotating single stars has been studied by
many investigators (Kippenhahn \& Thomas \cite{kippenhahn70}; Endal
\& Sofia \cite{endal76}; Pinsonneaul et al. \cite{pinsonneaul89};
Meynet \& Maeder \cite{meynet97}; Langer 1998, 1999; Huang 2004a).
However, it is also very important to study the evolution of
rotating binary stars (Jackson 1970; Chan \& Chau 1979; Langer 2003;
Huang 2004b; Petrovic et al 2005a,b; Yoon et al. 2006). The effect
of spin on structure equations has been investigated (e.g. the
present Eggleton's stellar evolution code; Li et al 2004a,b, 2005;
K$\ddot{a}$hler 2002). They adopted the lowest-order approximate
analysis in which two components were treated as spherical stars. In
fact, with the joint effects of spin and tide, the structure of a
star changes from spherically symmetric to non-spherically
symmetric. Then, the stellar structure equations become three
dimensional. Theory distinguishes two components in the tide, namely
equilibrium tide (Zahn 1966) and the dynamical tide (Zahn 1975).
Then, the dissipation mechanisms acting on those tides, namely the
viscous friction for the equilibrium tide and the radiative damping
for the dynamical tide, have been identified (Zahn 1966, 1975,
1977). The distortion throughout the outer regions of the two
components is not small in short-period binary systems. The
higher-order terms in the external gravitational field should not be
ignored (Jackson 1970).

It is a very complex process to determine the equilibrium structure
of the two components. Therefore, approximate methods have been
widely adopted for studying these effects.  In 1933, the theory of
distorted polytropes was introduced by Chandrasekhar. Kopal
(1972,1974) developed the concept of Roche equipotential and of
Roche coordinates to analyse the problem of rotationally and tidally
distorted stars in a binary system. Bur$\check{s}$a (1989a,1988)
took advantage of the high-order perturbing potential to describe
rotational and tidal deformations to discuss the figures and dynamic
parameters of synchronously orbiting satellites in the solar system.
The equilibrium structure of the two components were treated as two
non-symmetric rotational ellipsoids with two different semi-major
axes $a_{1}$ and $a_{2}$ ($a_{1}>a_{2}$) by Huang (2004b). It is
very important that Kippenhahn $\&$ Thomas (1970) introduced a
method of simplifying the two-dimensional model with conservative
rotation and allowed the structure equations for a one-dimensional
star to incorporate the hydrostatic effect of rotation. This method
has been adopted by Endal \& Sofia (1976) and Meynet $\&$ Maeder
(1997), who applied it to the case of shellular rotation law (Zahn
1992). In this case, the rotation rate takes the simplified form of
$\Omega=\Omega (r)$. It was demonstrated that the shape of an isobar
in the case of the shellular rotation law is identical to one of the
equipotentials in the conservative case of Meynet $\&$ Maeder
(1997).

At the semi-detached stage, both mass transfer between the
components and luminosity change of a secondary exist due to the
release of accretion energy which is correlative with the external
potential of the two components. When the joint effect of rotation
and tide are considered, the potential of the two components are
different from those in non-rotational cases. Therefore, the
luminosity due to the release of accretion energy, as well as
irradiation energy, can significantly alter the structure and
evolution of the secondary. In a rotating star, meridional
circulation and shear turbulence exist, both of which can drive the
transport of chemical elements. This effect is stronger and has
already been studied by many scholars (Endal $\&$ Sofia 1978;
Pinsonneaul et al. \cite{pinsonneaul89}; Chaboyer $\&$ Zahn 1992;
Zahn 1992; Meynet $\&$ Maeder 1997; Maeder 1997; Meader $\&$ Zahn
1998; Maeder $\&$ Meynet 2000; Denissenkov et al. 1999; Talon et al.
1997; Decressin et al.2009). In this paper, the amplitude expression
for the radial component of the meridional circulation velocity
$U(r)$ considers the effect of tidal force, which may be important
in a massive close binary system.

This paper is divided into four main sections. In section 2, the
structure equations of rotating binary stars are presented. Material
diffusion equations and boundary conditions are provided. Then, the
accretion luminosity, including gravitational energy, heat energy,
and radiation energy, is deduced. In section 3, the results of
numerical calculation are described and discussed in detail. In
section 4, conclusions are drawn.

\section{Model for rotating binary stars}

\subsection{Potential of rotating binary stars}

It is well known that the rotation of a component is synchronous
with the orbital motion of a system thanks to a strong tidal effect.
Such synchronous rotation also exists inside the component (Giuricin
et al. \cite{Giur84}; Van Hamme \& Wilson \cite {van90}); therefore,
conventional theories usually assume that two components rotate
synchronously and revolve in circular orbits (Kippenhahn \& Weigert
\cite{kippenhahn67}; De Loore \cite{de80}; Huang \& Taam
\cite{huang90}; Vanbeveren \cite{vanbeveren}; De Greve \cite
{de93}). A coordinate system rotating with the orbital angular
velocity of the stars is introduced. The mass centre of the primary
is regarded as the origin, and it is presumed that the z-axis is
perpendicular to the orbital plane, and the positive x-axis
penetrates the mass centre of the secondary. The gravitational
potential at any point $P(r,\theta,\varphi)$ of the surface of the
primary can be approximately expressed as
\begin{equation}
\Psi=V+\frac{1}{3}\Omega^{2}r^{2}(1-P_{2}(cos\theta))+V_{t},
\end{equation}
where $V$ is the gravitational potential and given by
Bur$\check{s}$e (1989a,1988),
\begin{equation}
V=\frac{GM_{1}}{r_{p}}[\frac{r_{p}}{r}+(\frac{r_{p}}{r})^{3}J_{2}^{(0)}P_{2}^{0}(\cos\theta)
+(\frac{r_{p}}{r})^{3}J_{2}^{(2)}P_{2}^{2}(\cos\theta)\cos
2\varphi].
\end{equation}
Here, $V_{t}$ is the tidal potential (Bur$\check{s}$e 1989a)
\begin{equation}
V_{t}=\frac{GM_{2}}{D}(\frac{r}{D})^{2}[-\frac{1}{2}P_{2}^{0}(\cos\theta)+\frac{1}{4}P_{2}^{2}(\cos\theta)cos
2\varphi],
\end{equation}
where it is assumed that the mean equatorial radius equals that of
the equivalent sphere in the above equation for the convenience of
calculation. Both $M_1$ and $M_2$ are the mass of the primary and
the secondary, respectively, and $r_{p}$ represents each equivalent
radius inside the star, $P_{2}^{2}(\cos\theta)$ and
$P_{2}^{0}(\cos\theta)$ are the associated Legendre
function($P_{2}^{0}(\cos\theta)=\frac{3}{2}\cos^{2}\theta-\frac{1}{2}$,
$P_{2}^{2}(\cos\theta)=3\sin^{2}\theta $), $D$ is the distance
between the two components, and $\Omega $ is the orbital angular
velocity of the system. It can be represented by
\begin{equation}
\Omega ^2=G(M_1+M_2)/D^3,
\end{equation}
where $J_{2}^{(0)}$ and $J_{2}^{(2)}$ are dimensionless stokes
parameters. If $M_{1}$ can generally be negligible compared to
$M_{2}$, the stokes parameters can be expressed as (Bur$\check{s}$e
1989a,1988)
\begin{equation}
J_{2}^{(0)}=-[\frac{1}{3}k_{s}+\frac{1}{2}k_{t}]q(\frac{r_{p}}{D})^{3}=-J_{2},
\end{equation}
\begin{equation}
J_{2}^{(2)}=\frac{1}{4}k_{t}q(\frac{r_{p}}{D})^{3},
\end{equation}
where $k_{s}$ is the secular Love number, which is expressed as a
measure of the body-yield-to-centrifugal deformation, and $k_{t}$ is
an analogous parameter that is introduced to describe the secular
tidal deformations. The response of the body to its centrifugal
acceleration and to the tidal perturbing potential is different in
the usual case. Therefore, the body-yield-to centrifugal deformation
is not equal to the body-yield-to-tidal deformation. If the subject
investigated is regarded as an ideal elastic body, the body-yield-to
centrifugal deformation is equal to the body-yield-to-tidal
deformation, $k_{s}=k_{t}$. In the ideal static equilibrium,
$k_{s}=k_{t}=1$ (Bur$\check{s}$e 1989a). We assume the ideal static
equilibrium in this paper. $q$ is the mass ratio of the secondary to
the primary ($q=\frac{M_{2}}{M_{1}}$). With Eqs. (2) and (3) being
combined with Eq. (1), the potential of the primary can be obtained
as
\begin{eqnarray}
\Psi_{P}&=&\frac{GM_{1}}{D}\{\frac{D}{r_{1}}+\frac{1}{2}
\frac{D}{r_{1}}\left(\frac{r_{p}}{r_{1}}\right)^{2}
[-J_{2}(3\cos^{2}\theta-1)\\
&& \nonumber+6J_{2}^{(2)}\sin^{2}\theta\cos2\varphi]
+\frac{1}{2}(1+q)(\frac{r_{1}}{D})^{2}\sin^{2}\theta
\\&&+\frac{1}{4}q(\frac{r_{1}}{D})^{2}[3\sin^{2}\theta(1+cos2\varphi)-2]\},
\nonumber
\end{eqnarray}
The potential of the secondary is deduced by substituting $M_{2}$
for $M_{1}$ and $\frac{1}{q}$ for $q$. The isobar defined by the
equation  $P=const$ is assumed to be a triaxial ellipsoid with three
semi-major axes: $a$, $b$, and $c$. The shortest axis defined by $c$
is identical to its rotational axis and perpendicular to its orbital
plane. The longest axis defined by $a$ is identical with its
$x$-axis.

\subsection{Considering stellar structure equations with spin and tidal effects}

The spin of the two components is rigid rotation, and it belongs to
conservative rotation. The definition of equivalent sphere was
adopted in a practical calculation. Therefore, the triaxial
ellipsoid model is simplified to a one-dimensional model. The
structure equations are presented as
\begin{equation}
\frac{\partial r_{P} }{\partial M_{P}}=\frac 1{4\pi r_{P}^2\rho},
\end{equation}
\begin{equation}
\frac{\partial P}{\partial M_{P} }=-\frac{GM_{P}}{4\pi r_{P} ^4}f_P,
\end{equation}
\begin{equation}
\frac{\partial L_{P}}{\partial M_{P} }=\varepsilon _N-\varepsilon
_\nu +\varepsilon _g +\sigma_{ac},
\end{equation}
where $\sigma_{ac}$  is the energy source per unit mass caused by
mass overflow and irradiation. Because accretion luminosity is
caused by energy sources in the gainer's outermost layer, there
exists
\begin{equation}
\sigma_{ac}\Delta m =\Delta L_{acc},
\end{equation}
where $\Delta m$ is the photosphere mass of the secondary. The
surface temperature of the secondary may be approximated by the
formula, $L_{2}+\Delta L_{acc}=4\pi R_{2}^{2}\sigma T_{eff}^{4}$,
where $L_{2}$ is the luminosity coming to the photosphere from the
stellar interior, and $\sigma$ is the Stefan-Boltzmann constant:
\begin{equation}
\frac{d\ln T}{d\ln P}=\Biggl\{\matrix {\nabla
_{\mathrm{R}}f_{\mathrm{T}}/ f_{\mathrm{p}}\cr \nabla
_{\mathrm{con}}\cr }
\end{equation}
\begin{equation}
f_P=\frac{4\pi r_{p}^4}{GM_{p} S_{p} }\frac 1{<g_{eff}^{-1}>} ,
\end{equation}
\begin{equation}
f_T=\frac{4\pi r_{p} ^2}{S_{p}} \frac {1}{<g_{eff}><g_{eff}^{-1}>} ,
\end{equation}
\begin{equation}
\nabla _R=\frac{3}{16\pi acG}\frac{\kappa LP}{M_{P} T^4},
\end{equation}
where $<g_{eff}> and <g_{eff}^{-1}>$ are the mean values of
effective gravity and its opposites over the isobar surface, and
$\nabla _R$ is the radiative temperature gradient.  The factors
$f_P$ and $f_T$ depend on the shape of the isobars.

\subsection{ Calculation of quantities $f_P$ and $f_T$}
\subsubsection{Shape and gravitational acceleration of triaxial ellipsoid}

To obtain the factors $f_P$ and $f_T$, the mean values $<g_{eff}>$
and $<g_{eff}^{-1}>$ over the isobar surface have to be calculated.
Therefore, the shape of isobars must be given first. The functions
for the semi-major axes $a$, $b$, and $c$ to the radius of the
equivalent sphere $r_{P} $ can be obtained from Eq. (7) as
\begin{eqnarray}
\frac{4\pi abc}3 =\frac{4\pi r_{P} ^3}3 ,
\end{eqnarray}
\begin{eqnarray}
&\frac{GM_1}D[\frac{D}{a}+\frac 12
\frac{D}{a}(\frac{r_{p}}{a})^{2}(J_{2}+6J_{2}^{(2)}))+
\\
& \nonumber\frac{1}{2}(1+q)(\frac{a}{D})^{2}+q(\frac{a}{D})^{2}]
=\frac{GM_1}D[ \frac{D}{c}+\frac 12
\frac{D}{c}(\frac{r_{p}}{c})^{2}(-2J_{2})\\
&-\frac{1}{2}q(\frac{c}{D})^{2}], \nonumber
\end{eqnarray}
\begin{eqnarray}
&\frac{GM_1}{D}[\frac{D}{b}+\frac{1}{2}
\frac{D}{b}(\frac{r_{p}}{b})^{2}(J_{2}-6J_{2}^{(2)})+\frac{1}{2}(1+q)
(\frac{b}{D})^{2}\\&\nonumber-\frac{1}{2}q(\frac{b}{D})^{2}]=\frac{GM_1}D[
\frac{D}{c}+\frac 12 \frac{D}{c}(\frac{r_{p}}{c})^{2}(-2J_{2})\\&
\nonumber-\frac{1}{2}q(\frac{c}{D})^{2}]. &\nonumber
\end{eqnarray}
The left hand side of Eq.(17) corresponds to $\theta=\frac{\pi}{2}$
and $\varphi=0$, while the one of Eq.(18) corresponds to
$\theta=\frac{\pi}{2}$ and $\varphi=\frac{\pi}{2}$. The three
semi-major axes $a$, $b$, and $c$ of a triaxial ellipsoid can be
obtained numerically by solving (16), (17), and (18). From (7), the
quantities $g_{r}$, $g_{\theta}$, and $g_{\varphi}$ at the surface
of the two components take the forms of
\begin{eqnarray}
&g_{r}=-\frac{\partial \Psi}{\partial
r}=\frac{GM_{1}}{D^{2}}\{(\frac{D
}{r})^{2}+\frac{3}{2}\frac{D^{2}}{r^{2}}(\frac{r_{p}}{r})^{2}\\
&\nonumber[-J_{2}(3\cos^{2}\theta-1)+6J_{2}^{(2)}\sin^{2}\theta\cos2\varphi]
-\frac{r}{D}(1+q)\sin^{2}\theta\\
&\nonumber-\frac{r}{D}q(3\sin^{2}\theta\cos^{2}\varphi-1)\},&
\end{eqnarray}
\begin{eqnarray}
&g_{\theta}=-\frac{1}{r}\frac{\partial \Psi}{\partial
\theta}=\frac{GM_{1}}{D^{2}}\{-\frac{D^{2}}{r^{2}}
(\frac{r_{p}}{r})^{2}[3J_{2}+6J_{2}^{(2)}\\
&\nonumber(2\cos^{2}\varphi-1)]-(1+q)\frac{r}{D}
-3q\frac{r}{D}\cos^{2}\varphi \}\sin\theta \cos\theta &,\nonumber
\end{eqnarray}
\begin{eqnarray}
&g_{\varphi}=-\frac{1}{r\sin \theta}\frac{\partial \Psi}{\partial
\varphi}\\&\nonumber=\frac{GM_{1}}{D^{2}}[12(\frac{D}{r})^{2}(\frac{r_{p}}{r})^{2}J_{2}^{(2)}
+3q\frac{r}{D}]\sin\theta \cos\varphi \sin\varphi.
\end{eqnarray}
However, the total potential in the stellar interior (to first-order
approximation) can be composed by four parts (Kopal 1959, 1960,
1974; Endal \& Sofia 1976 and Landin 2009): $\psi_{s}$, the
spherical symmetric part of the gravitational potential; $\psi_{r}$,
the cylindrically symmetric potential due to rotation; $\psi_{t}$
the non-symmetric potential due to tidal force, and $\psi_{d}$, the
non-symmetric part of the gravitational potential due to the
distortion of the component considering the rotational and tidal
effects. Therefore, the total potential at $P(r,\theta,\varphi)$ is
\begin{eqnarray}
\Psi&=&\psi_{s}+\psi_{r}+\psi_{t}+\psi_{d}\nonumber\\&&=
\frac{GM_{\psi}}{r}+\frac{1}{2}\omega^{2}r^{2}\sin^{2}\theta
+\frac{GM_{2}}{D}[1+\sum_{j=2}^{4}(\frac{r_{0}}{D})^{j}P_{j}(\sin\theta\cos\varphi)]\nonumber\\
&&-\frac{4\pi}{3r^{3}}P_{2}(\cos\theta)\int_{0}^{r_{0}}\rho\frac{
r_{0}'^{7}}{M_{\psi}}\Omega^{2}\frac{5+\eta_{2}}{2+\eta_{2}}dr_{0}'\\
&&+4\pi
GM_{2}\sum_{j=2}^{4}\frac{P_{j}(\sin\theta\cos\varphi)}{(rD)^{j+1}}\int_{0}^{r_{0}}\rho\frac{
r_{0}'^{2j+3}}{M_{\psi}}\frac{j+3+\eta_{j}}{j+\eta_{j}}dr_{0}'.
\nonumber
\end{eqnarray}
The quantity $\eta_{j}$ can be evaluated by numerically integrating
the Radau's equation (cf. Kopal 1959)
\begin{equation}
r_{0}\frac{d\eta_{j}}{dr_{0}}+6\frac{\rho(r_{0})}{\overline{\rho}(r_{0})}
(\eta_{j}+1)+\eta_{j}(\eta_{j}-1)=j(j+1),
\end{equation}
for j=2,3,4, and boundary condition $\eta_{j}(0)=j-2$. The quantity
$r_{0}$ is the mean radius of the corresponding isobar. The local
effective gravity is given by differentiation of the total potential
and is written as
\begin{equation}
g_{i}=[(\frac{\partial \psi}{\partial
r})^{2}+(\frac{1}{r}\frac{\partial \psi}{\partial
\theta})^{2}+(\frac{1}{r\sin\theta}\frac{\partial \psi}{\partial
\varphi})^{2}]^{\frac{1}{2}}, (i=1,2).
\end{equation}
The integral in above equations and their derivatives must be
evaluated numerically. The mean values of $g_{effi}$ and
$g_{effi}^{-1}$ over the surfaces of the triaxial ellipsoids can be
obtained as
\begin{equation}
<g_{effi}>=\frac 1{S_{P}}\int_0^{\pi}\int_0^{2 \pi}g_{effi}
r_{i}^2\sin \theta d\theta d \varphi, (i=1,2)
\end{equation}
\begin{equation}
<g_{effi}^{-1}>=\frac 1{S_{P}}\int_0^{\pi}\int_0^{2
\pi}g_{effi}^{-1} r_{i}^2\sin \theta d\theta d \varphi, (i=1,2).
\end{equation}
According to Eqs. (13) and (14), the values of $f_P$ and $f_T$ can
be obtained when the mean values $<g_{eff}>$ and $<g_{eff}^{-1}>$
are known. $r_{1}$ and $r_{2}$  are the distances between the centre
of the components and the surfaces of two triaxial ellipsoids. They
are
\begin{eqnarray}
r_{i}^2=\frac{a_{i}^{2}b_{i}^{2}c_{i}^{2}}{b_{i}^{2} c_{i}^{2}
\sin^{2}\theta \cos^{2}\varphi+a_{i}^{2} c_{i}^{2}\sin^{2}\theta
\sin^{2}\varphi+a_{i}^{2}b_{i}^{2}
cos^{2}\theta}\\&\nonumber,(i=1,2).
\end{eqnarray}
The surface area $S_{p}$ of the isobar can be expressed
as
\begin{equation}
S_{p} =\frac{4\pi}{3}(a^{2}+b^{2}+c^{2}) .
\end{equation}

\subsection{Element diffusion process}

The effect of meridian circulation can drive the transport of
chemical elements and angular momentum in rotating stars. For the
components in solid-body rotation, no differential rotation exists
that can cause shear turbulence. According to Endal \& Sofia (1978)
and Pinsonneault (1989), the transport of chemical composition is
treated as a diffusion process. The equation takes the form of
(Chaboyer \& Zahn 1992)
\begin{eqnarray}
\left( \frac{\partial y_\alpha }{\partial t}\right) &=&\frac{1}{\rho
r^{2}} \frac\partial {\partial r}\left[\rho r^{2} D_{dif}\left(
\frac{\partial y_\alpha }{\partial r}\right)\right] +\left(
\frac{\partial y_\alpha }{\partial t}\right) _{nuc},
\end{eqnarray}
where $(\frac{\partial y_\alpha }{\partial t})_{nuc}$ is a source
term from nuclear reactions, and $y_{\alpha}$ is the relative
abundance of $\alpha-th$ nuclide. The diffusion coefficient
$D_{dif}$ given by Heger, Langer \& Woosley (2000a) can be expressed
as
\begin{equation}
D_{dif}\equiv min\{d_{inst}, H_{v,ES}\}U(r),
\end{equation}
where $d_{inst}$ and $H_{v,ES}$ denote the extent of the instability
and the velocity scale height, respectively. The expression for the
amplitude of the radial component of the meridional circulation
velocity $U(r)$ (derived from  Kippenhahn $\&$ Weight 1990) has been
modified to take the effects of radiation pressure and tidal force
into-account,  which are important in a massive close binary system.
It is noticed that
\begin{eqnarray}
&U(r)=\frac{8}{3}(\frac{\Omega^{2}r^{3}}{GM_{r}}+\frac{GM_{2}r^{3}}{GM_{r}D^{^{3}}})
\frac{L_{r}}{g_{r}M_{r}}\frac{\gamma-1}{\gamma}\frac{1}
{\nabla_{ad}-\nabla}&\\& \nonumber (1-\frac{\Omega^{2}}{2\pi
G\rho}-\frac{\varepsilon}{\varepsilon_{m}})&
\end{eqnarray}
where the term
$\frac{\Omega^{2}r^{3}}{GM_{r}}+\frac{M_{2}r^{3}}{M_{r}D^{^{3}}}$ is
the local ratio of centrifugal force and tidal force to gravity,
$\gamma$ is the ratio of the specific heats $\frac{C_{p}}{C_{v}}$,
$L_{r}$ represents the luminosity at radius $r$, $M_{r}$ is the mass
enclosed within a sphere of radius r, $\nabla$ is the actual
gradient and $\nabla_{ad}$ is adiabatic temperature gradient,
$\varepsilon_{m}=\frac{L}{m}$ gives the mean energy production rate,
and  $\varepsilon$ is local generation rate of nuclear-energy.

There is no source or sink at the inner and the outer boundaries of
the two components. Therefore, the boundary conditions are used as
\begin{equation}
(\frac{\partial y_\alpha}{\partial r})_{i=1}=0=(\frac{\partial
y_\alpha}{\partial r})_{i=M}
\end{equation}
where the subscript $i$ denotes different layers inside stars. The
initial abundance equals the one at the zero-age main sequence.
Therefore, the initial condition is
\begin{equation}
(y_\alpha)_{i}|_{t=0}=(y_\alpha)_{i}|_{init}.
\end{equation}

\subsection{Luminosity accretion}

In the case where the joint effect of rotation and tide is ignored,
the two components are spherically symmetric. The star fills its
Roche lobe and begins to transfer matter to the companion. However,
in the case with the effects of rotation and tide being considered,
the components are triaxial ellipsoids. The condition for the mass
overflow through Roche lobe flow should be revised as $a_1=r_{robe}$
(Huang 2004b). It is assumed that the transferred mass is
distributed within a thin shell at the surface of the primary before
the transfer, and within a thin shell at the surface of the
secondary after the transfer. Three forms of energy (including
potential energy, heat energy, and radiative energy) are transferred
to the secondary. The mass transfer rate is $\dot{m}$. Two different
cases are considered:

a) If the joint effect of rotation and tide is ignored, the
accretion luminosity can be expressed directly in terms of the Roche
lobe potential at the inner Lagrangian point, $\Psi_{L_{1}}$, and at
the surface of the secondary $\Psi_{s}$ (Han \& Webbink 1999):
\begin{eqnarray}
&\triangle L_{P}=
\dot{m}(\Psi_{L_{1}}-\Psi_{S})=\frac{G\dot{m}M_1}{D}
[\frac{1}{X_{L_{1}}/D}+\frac{q}{1-X_{L_{1}}/D}+\frac{1+q}{2}(\frac{X_{L_{1}}}{D}\nonumber\\
&-\frac{q}{1+q})^{2}-\frac{q}{R_{2}/D}-\frac{1}{1-R_{2}/D}
-\frac{1+q}{2}(1-\frac{R_{2}}{D}-\frac{q}{1+q})^{2})],&
\end{eqnarray}
where $X_{L1}$ is the distance between the primary and $L_{1}$, and
$R_{2}$  is the radius of the secondary.

b) If the joint effect of rotation and tide is considered, the
equilibrium structure of the two components will be treated as
triaxial ellipsoids. The release of potential energy because of the
accretion of a mass rate $\dot{m}$ to the secondary is given by
\begin{eqnarray}
&\triangle L_{P}=
\dot{m}(\Psi_{L_{1}}-\Psi_{S})=\frac{G\dot{m}M_1}{D}[\frac{1}
{X_{L_{1}}/D}+\frac{q}{1-X_{L_{1}}/D}+\frac{1+q}{2}(\frac{X_{L_{1}}}{D}\nonumber\\
&-\frac{q}{1+q})^{2}]-\frac{GM_2\dot{m}}{D}[ \frac{ D}{c_{2}}+\frac
12\frac{D}{c_{2}}(\frac{r_{p}}{c_{2}})^{2}(2J_{2}^{0})-\frac{1}{2}q
(\frac{c_{2}}{D})^{2}],
\end{eqnarray}
where $\Psi_{s}$ is the potential of the secondary. Similarly, as
the two components have different temperatures, the transmitted
thermal energy will be
\begin{equation}
\triangle
L_{T}=\dot{m}(\frac{3kT_{eff1}}{2\mu_{1}m_{p}}-\frac{3kT_{eff2}}{2\mu_{2}m_{p}}),
\end{equation}
where $T_{eff1}$ and $T_{eff2}$ represent the effective temperature
of the primary and the secondary, respectively, and $\mu_{1}$ and
$\mu_{2}$ are the mean molecular weights of the primary and the
secondary, respectively.  $m_{p}$ refers to proton mass. Because of
the irradiation, energy accumulated by the primary and the secondary
can be given by (Huang \& Taam 1990)
\begin{equation}
\triangle L_{r,1,2}=
\frac{1}{2}[1-(D^{2}-R_{1,2}^{2})^{\frac{1}{2}}/D]L_{2,1},
\end{equation}
where $R_{1}$ and $R_{2}$ are the radii of the primary and the
secondary, and $L_{1}$ and $L_{2}$ are the luminosities of the
primary and the secondary, respectively. The total accretion
luminosity is
\begin{equation}
\triangle L_{acc}=\beta \triangle L_{tot}=\beta(\Delta L_{P}+\Delta
L_{T}+\Delta L_{r}).
\end{equation}
Because a part of the total energy may be dissipated dynamically,
$\beta$ is assumed to range from $0.1$ to  $0.5$ (Huang 1993). A
value $\beta=0.3$  is adopted.

\section{Results of numerical calculation}

The structure and evolution of binary system was traced with the
modified version of a stellar structure program, which was developed
by Kippenhahn et al., (1967) and has been updated to include mass
and energy transfer processes. The calculation method is based on
the technique of Kippenhahn and Thomas (1970) and takes advantage of
the concept of isobar (Zahn 1992, Meynet and Maeder 1997). Both
components of the binary are calculated simultaneously. The initial
mass of the system components is set at 9$M_{\odot}$ and
6$M_{\odot}$. The initial chemical composition X equals X=0.70, and
Z=0.02 is adopted for the two components. Similarly, the initial
orbital separation between the two components for all sequences is
defined as 20.771$R_{\odot}$, so mass transfer via Roche lobe occurs
in case A (at the central hydrogen-burning phase of the primary).
Two evolutionary sequences corresponding to the evolution with the
joint effect of rotation and tide being considered or ignored are
calculated. The sequence denoted by case 1 represents the evolution
without the effects of rotation and tide being considered, while the
sequence denoted by case 2 represents the evolution with the effects
of rotation and tide being considered. The calculation of Roche lobe
is taken from the study by Huang \& Taam (1990).  The
non-conservative evolution in the two cases was considered. Because
the local flux at colatitude $\theta$ is proportional to the
effective gravity $g_{e}$ according to Von Zeipel theorem (Maeder
1999), the mass-loss rate due to the stellar winds intensified by
tidal, rotational, and irradiative effects is obtained according to
Huang \& Taam(1990) (cf.Table 1). The angular velocity of the system
and the orbital separation between the two components change due to
a number of factors: changes in physical processes as the binary
system evolves, including the loss of mass and angular momentum via
stellar winds, mass transfer via Roche lobe overflow, exchange of
angular momentum between component rotation and the orbital motion
of the system caused by tidal effect, and changes in moments of
inertia of the components. The changes in the angular velocity of
the system and the orbital separation between the two components can
be calculated according to Huang \& Taam (1990), and the results are
listed in Table 1. Other parameters are treated in the same way for
two sequences.

\begin{table*}[tbp]
\caption{Parameters at different evolutionary points a, b, c, d, e,
and f in sequences of cases 1 and 2. } \label{table1}
\begin{flushleft}
\begin{tabular}{cccccccccccccccc} \hline\hline   Sequence &
$Time $ & $P$ & $M _{1} $ & $M _{2} $ & $log L_{1}/L_{\odot}$ & $log
T_{1,eff}$ &$log L_{2}/L_{\odot}$ & $log T_{2,eff}$ & $Y_{1}(c)$ &
$Y_{1}$ &
$V_{rot,1}$ & $V_{rot,2}$\\

 & $10^{7}yr$ & $day$ & $M_{\odot}$ & $M_{\odot}$ & $$ & $$ & $$ &
$$ & & $$ & $km/sec$ & $km/sec$\\

\hline

 a  &        &       &       &       &       &       &        &        &         &        &      &   \\
Case 1 & 0.0000 & 2.777 & 9.000 & 6.000 & 3.639 & 4.385 &3.077& 4.287& 0.2800  & 0.2800 &     &    \\
Case 2& 0.0000 & 2.776 & 9.000 & 6.000 & 3.629 & 4.381 &3.055  &
4.280  & 0.2800 & 0.2800 & 68.66  & 56.39\\ \hline
 b  &        &       &       &       &       &       &       &        &         &        &       &  \\
Case 1& 2.6725 & 2.737& 8.929& 5.994  & 3.959 & 4.285 &  3.110 & 4.267 & 0.8744 & 0.2800 &        & \\
Case 2& 2.6267 & 2.743 & 8.939 & 5.995  &  3.914& 4.295 &  3.085&
 4.265& 0.8247 & 0.2800 & 143.69 & 63.39\\ \hline
 c  &        &       &       &       &       &       &       &        &         &        &       &  \\
Case 1& 2.6854& 3.190 & 5.314& 9.608& 3.457&  4.120 & 3.811& 4.396 &  0.8799&  0.2801 &       &  \\
Case 2& 2.6329 & 3.192 &5.318& 9.616 & 3.264& 4.135 &3.834& 4.406&
0.8267& 0.2801 & 121.81 &  67.59\\ \hline
 d  &        &       &       &       &       &       &       &        &         &        &        & \\
Case 1& 3.0784 & 11.743& 2.727 & 12.168& 3.776&4.128 & 4.145 &4.461& 0.9800 &0.5892 &       &  \\
Case 2& 3.3567 & 7.213 & 3.388 & 11.497& 3.442 &4.162& 4.123&4.405
&0.9800 & 0.3481 &58.41& 41.79\\ \hline

e  &        &       &       &       &       &       &       &        &         &        &        & \\
Case 1& 3.0941 & 32.534& 1.797 & 13.082& 3.959& 4.028 & 4.276&4.539 &  0.9799 &0.8775 &       &  \\
Case 2    & &  & & &  &&  &   & &  & & \\ \hline

f  &        &       &       &       &       &       &       &        &         &        &        & \\
Case 1& 3.1484& 42.382& 1.572 & 13.245&3.186 &4.775 & 4.308& 4.549 & 0.8856& 0.8793&        & \\
Case 2& 3.5470& 32.531& 1.714& 13.037& 3.280& 4.686& 4.308&
4.419&0.9800 & 0.8261 & 0.97 &10.74 \\ \hline\hline

\end{tabular}
\end{flushleft}
\end{table*}

\begin{figure*}
\centering
\includegraphics[width=7.5cm,clip]{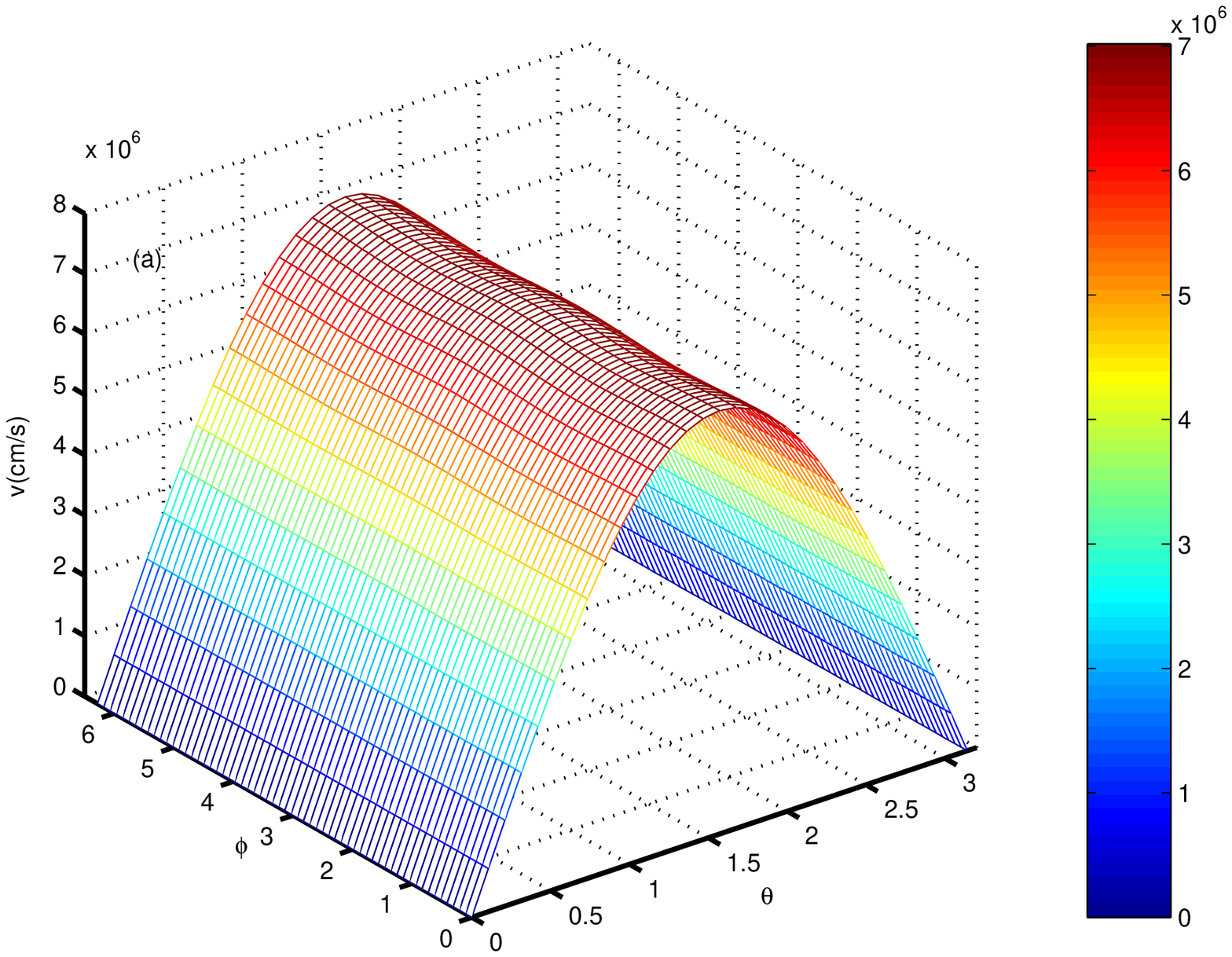}
\includegraphics[width=7.5cm,clip]{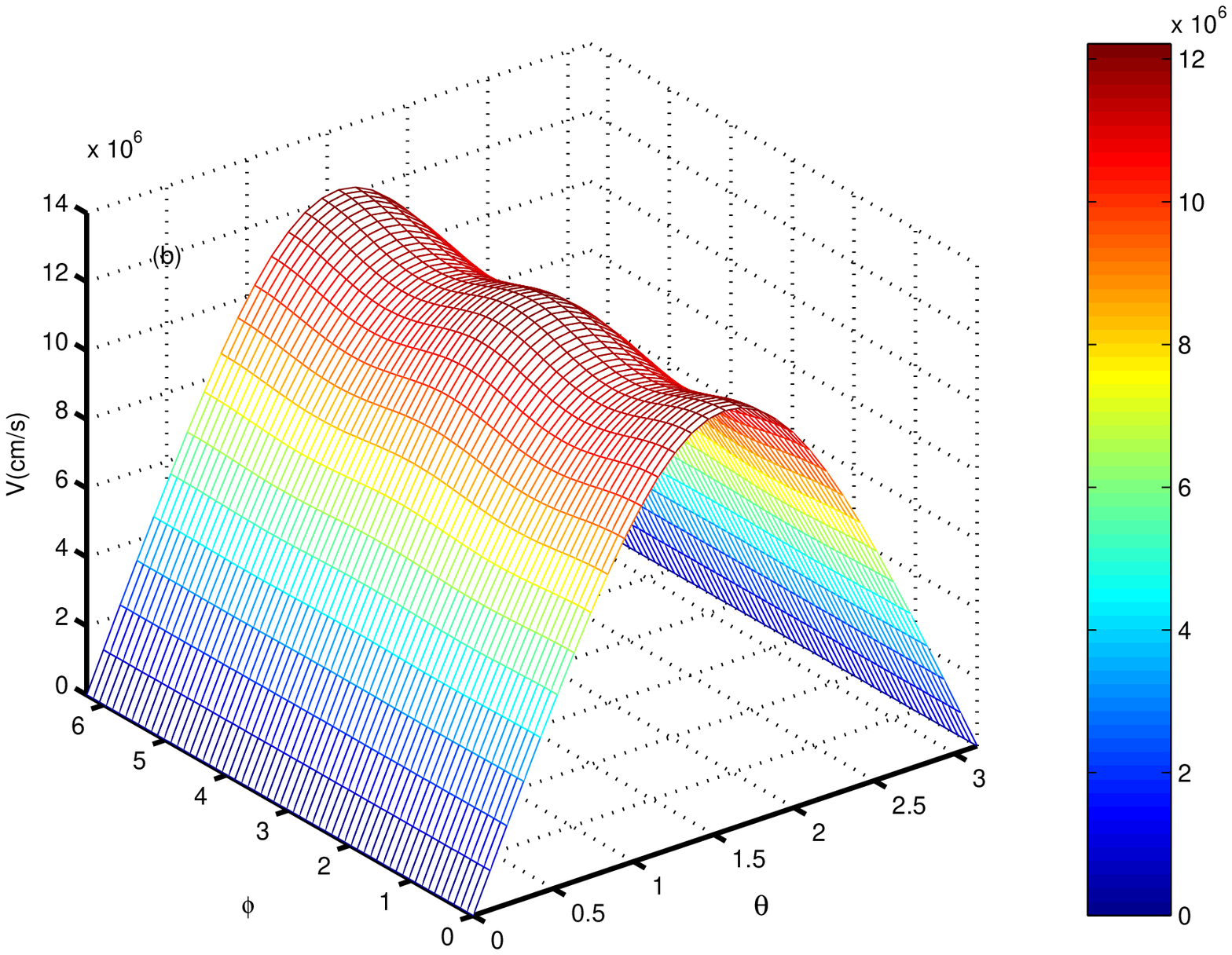}
\includegraphics[width=7.5cm,clip]{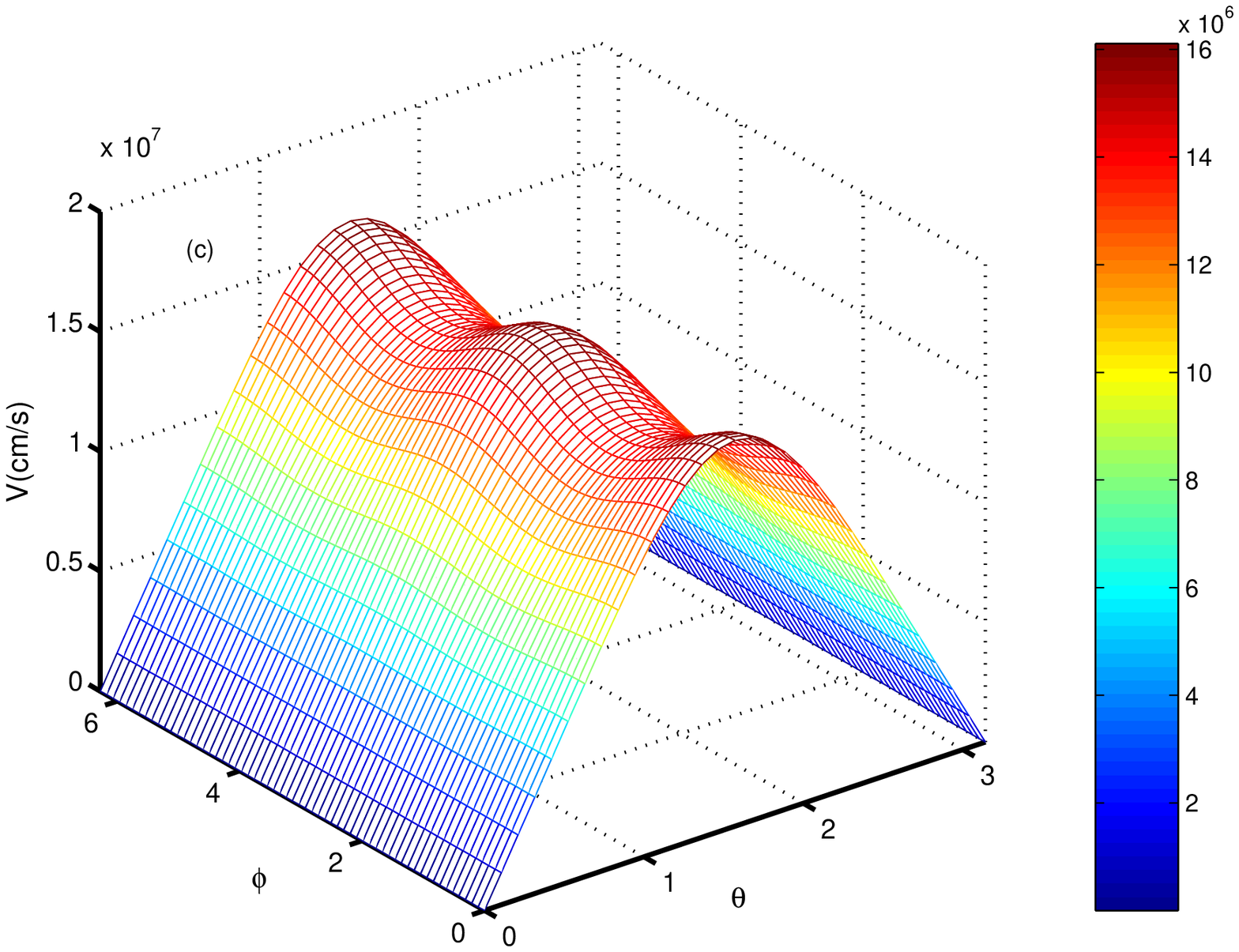}
\includegraphics[width=7.5cm,clip]{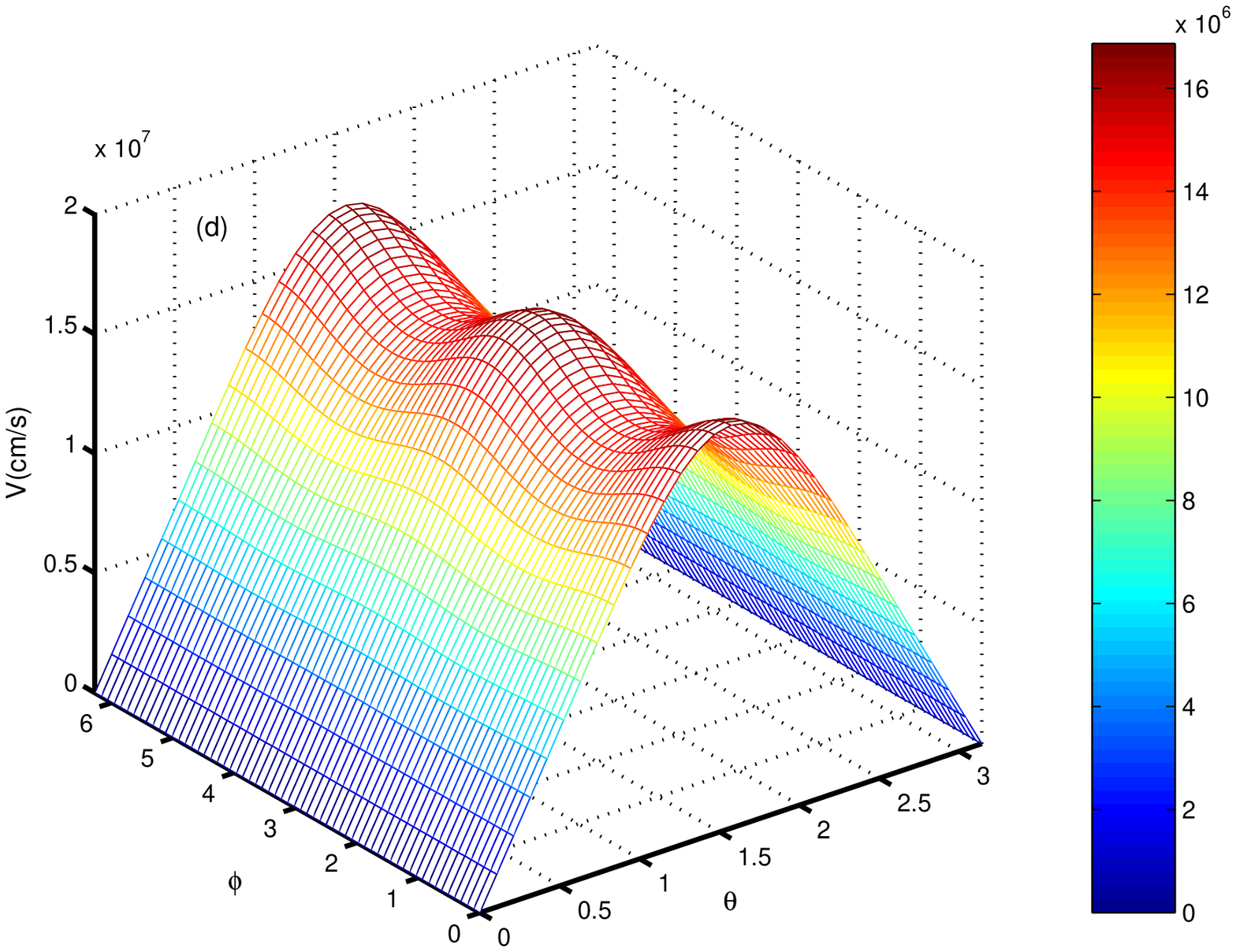}
\caption{Surface rotating velocity distribution of primary varying
with time. Four panels (a), (b), (c), and (d) correspond to periods:
$2.776$, $2.760$, $2.746$, and $2.628$ days, and corresponding
evolutive time is 0, $2.3386\times 10^{7}$, $2.6194\times 10^{7}$,
$2.6287\times 10^{7}$ yrs, respectively. } \label{appfig}
\end{figure*}
\begin{figure*}
\centering
\includegraphics[width=7.5cm,clip]{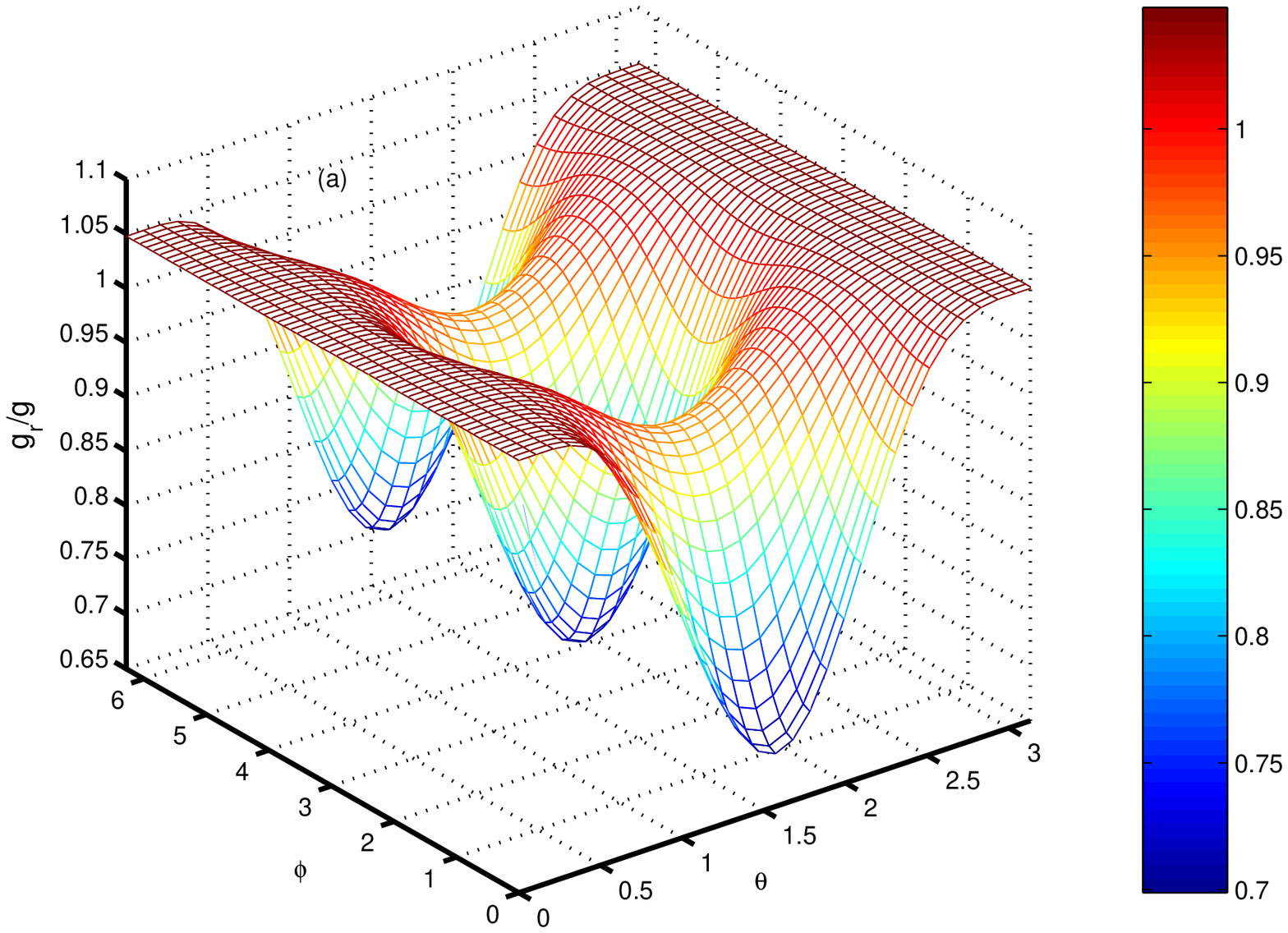}
\includegraphics[width=7.5cm,clip]{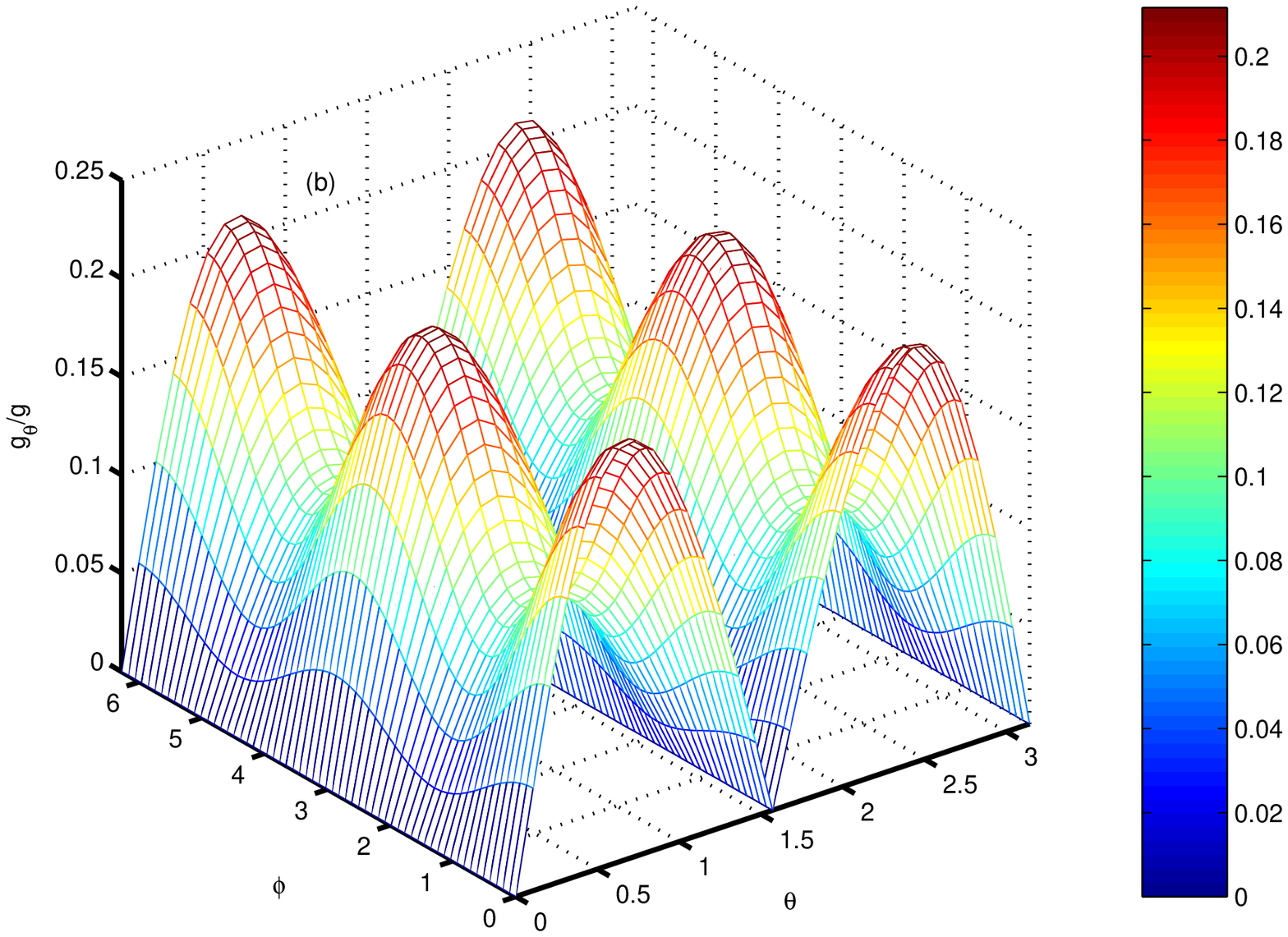}
\includegraphics[width=7.5cm,clip]{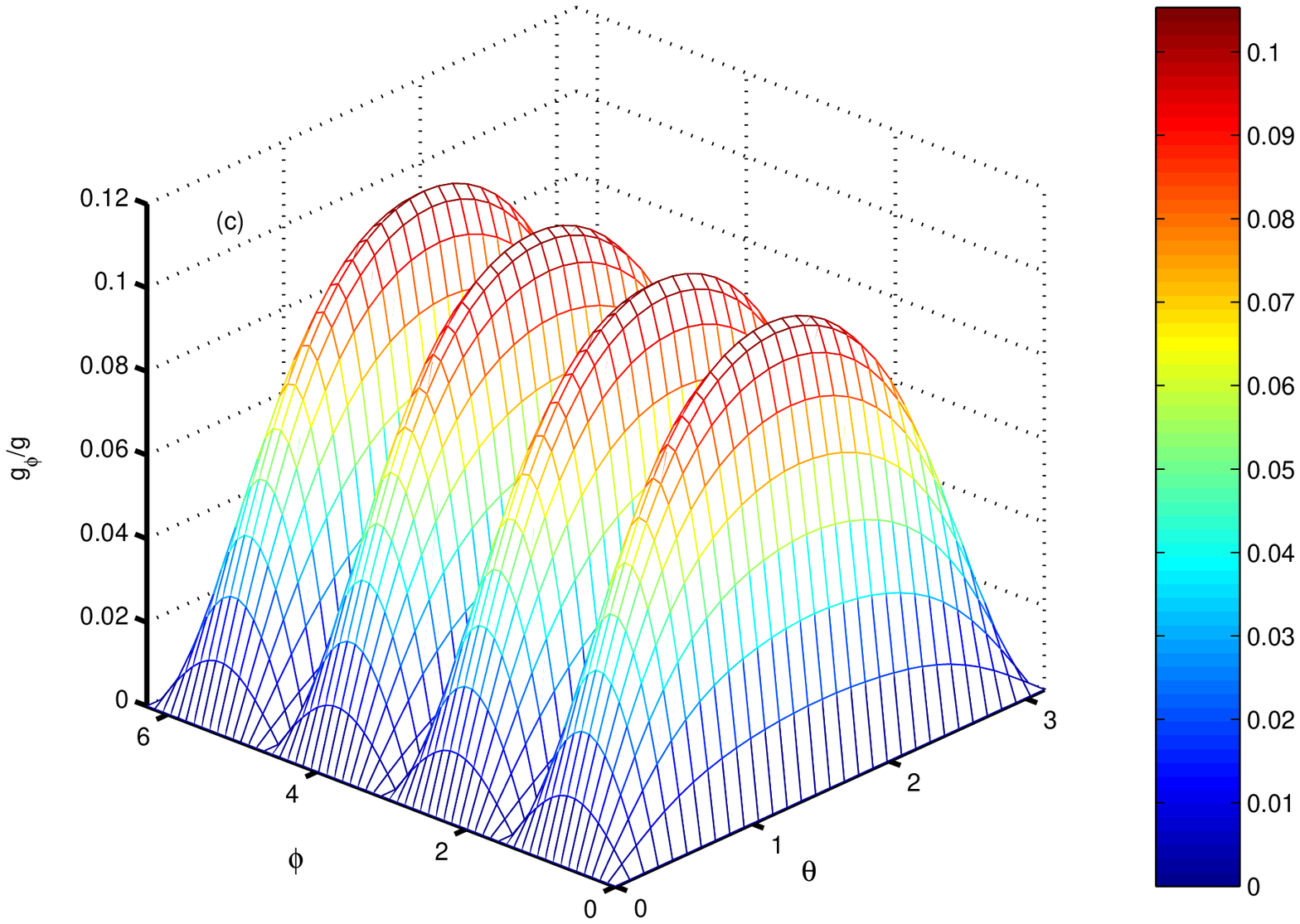}
\includegraphics[width=7.5cm,clip]{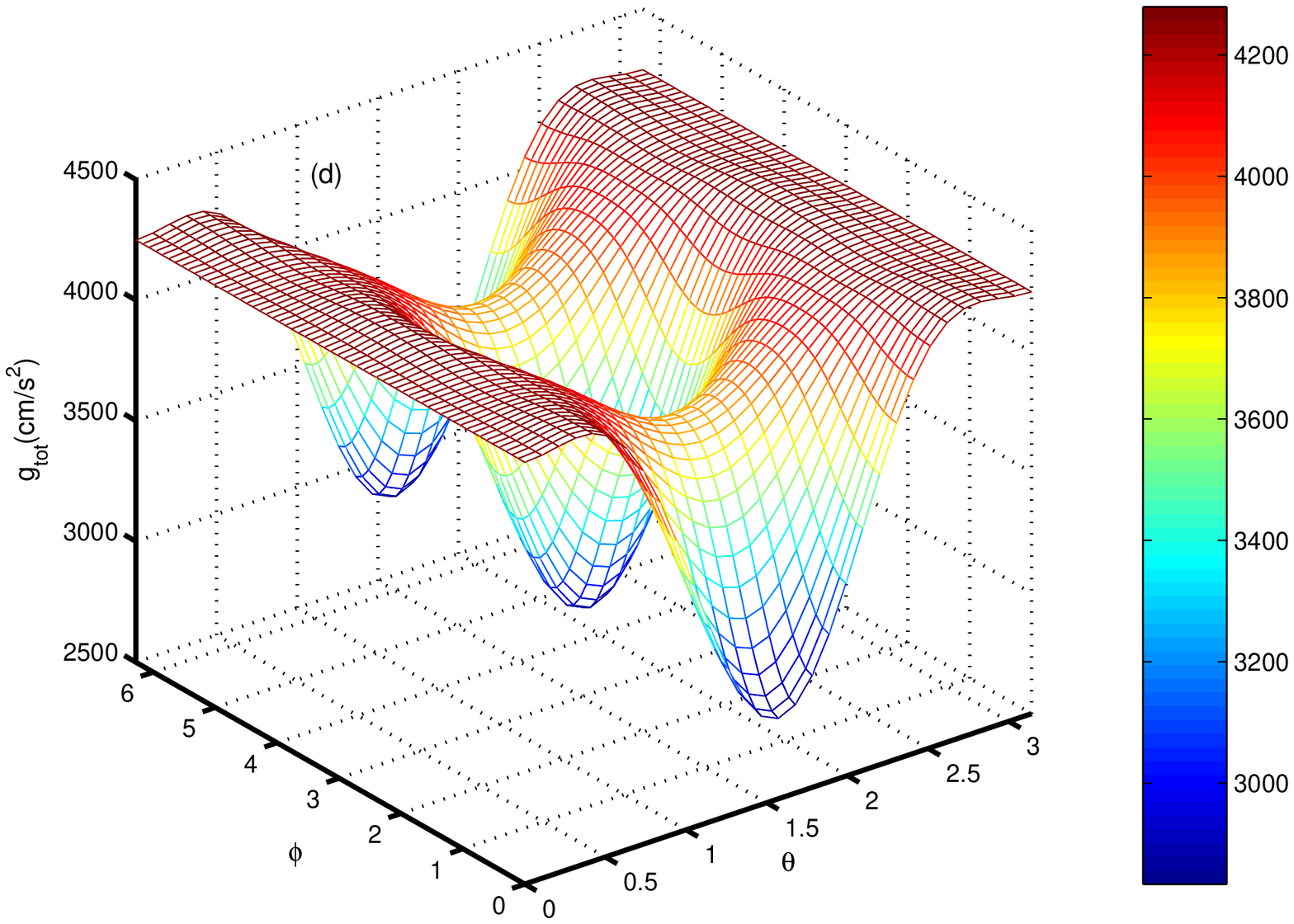}
\includegraphics[width=7.5cm,clip]{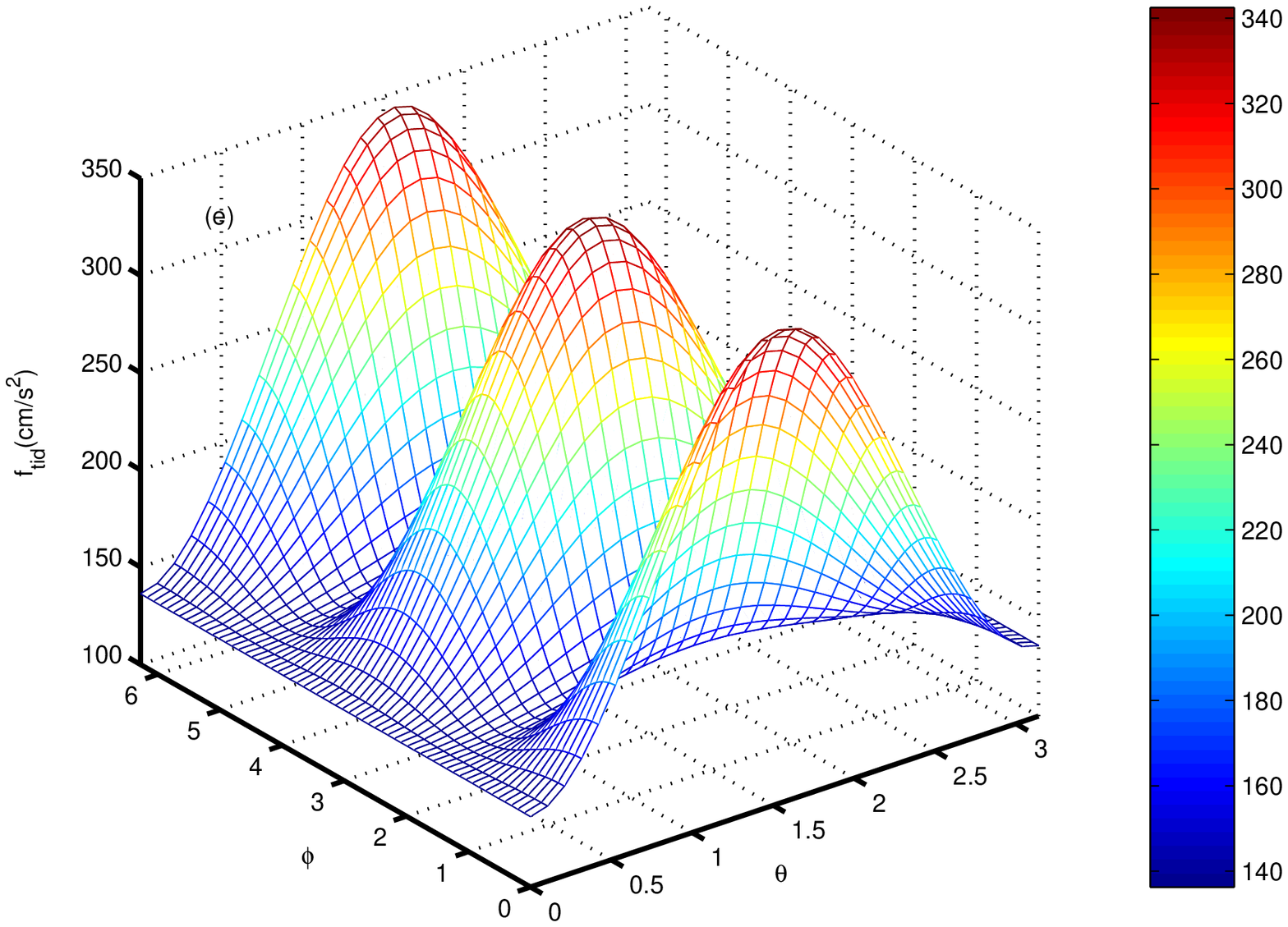}
\includegraphics[width=7.5cm,clip]{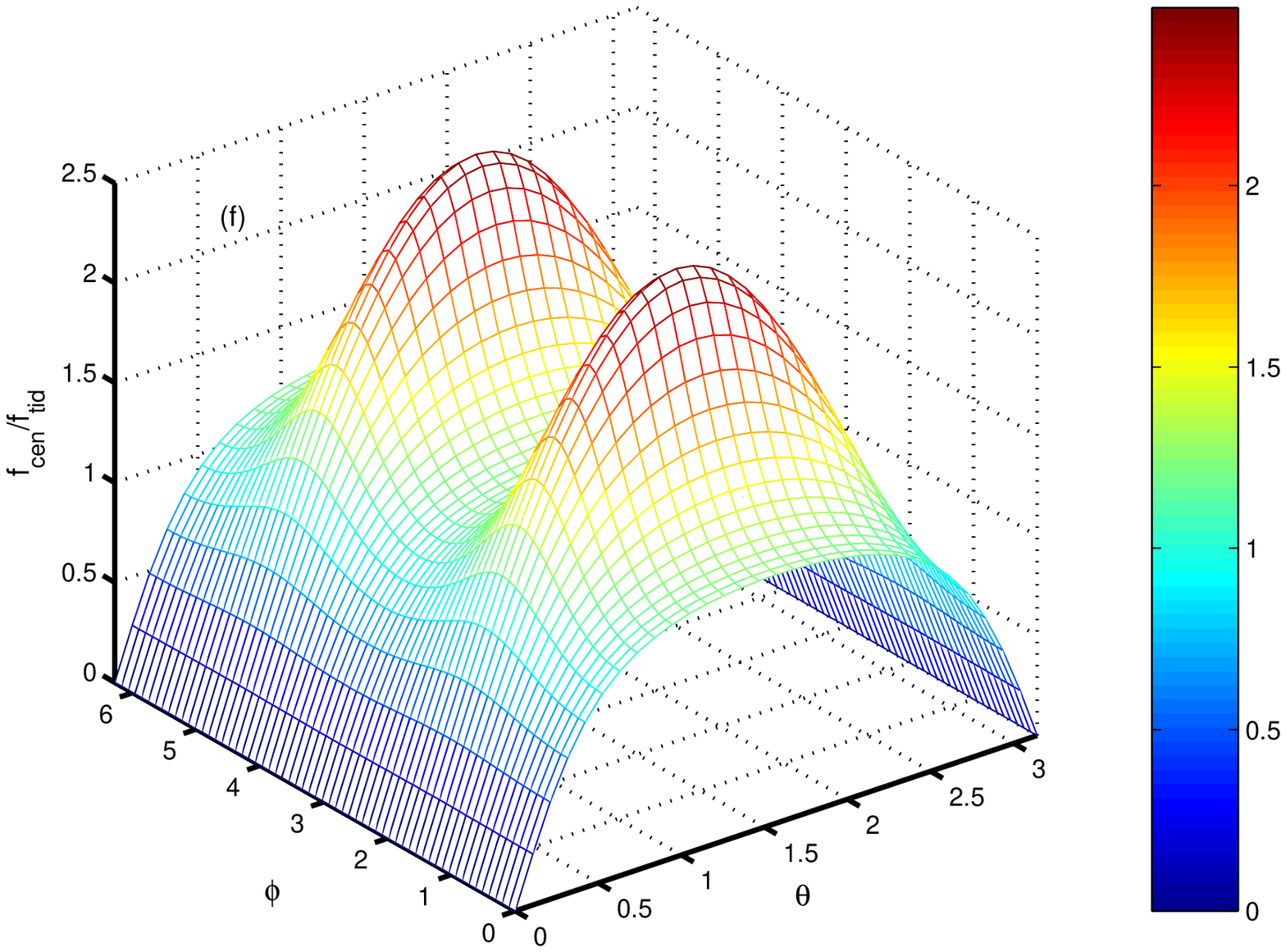}
\caption{Variation of relative gravitational accelerations at the
surface of primary under coordinate $\theta$ and $\varphi$ as mass
overflow begins. The quantities $g_{r}$, $g_{\theta}$, and
$g_{\varphi}$ are the three components of the gravitational
acceleration $g_{tot}$. Quantity $g$ equals the gravitational
acceleration of the corresponding equivalent sphere
($g=\frac{GM_{1}}{r_{p}^{2}}$).} \label{appfig}
\end{figure*}
\begin{figure*}
\centering
\includegraphics[width=7.5cm,clip]{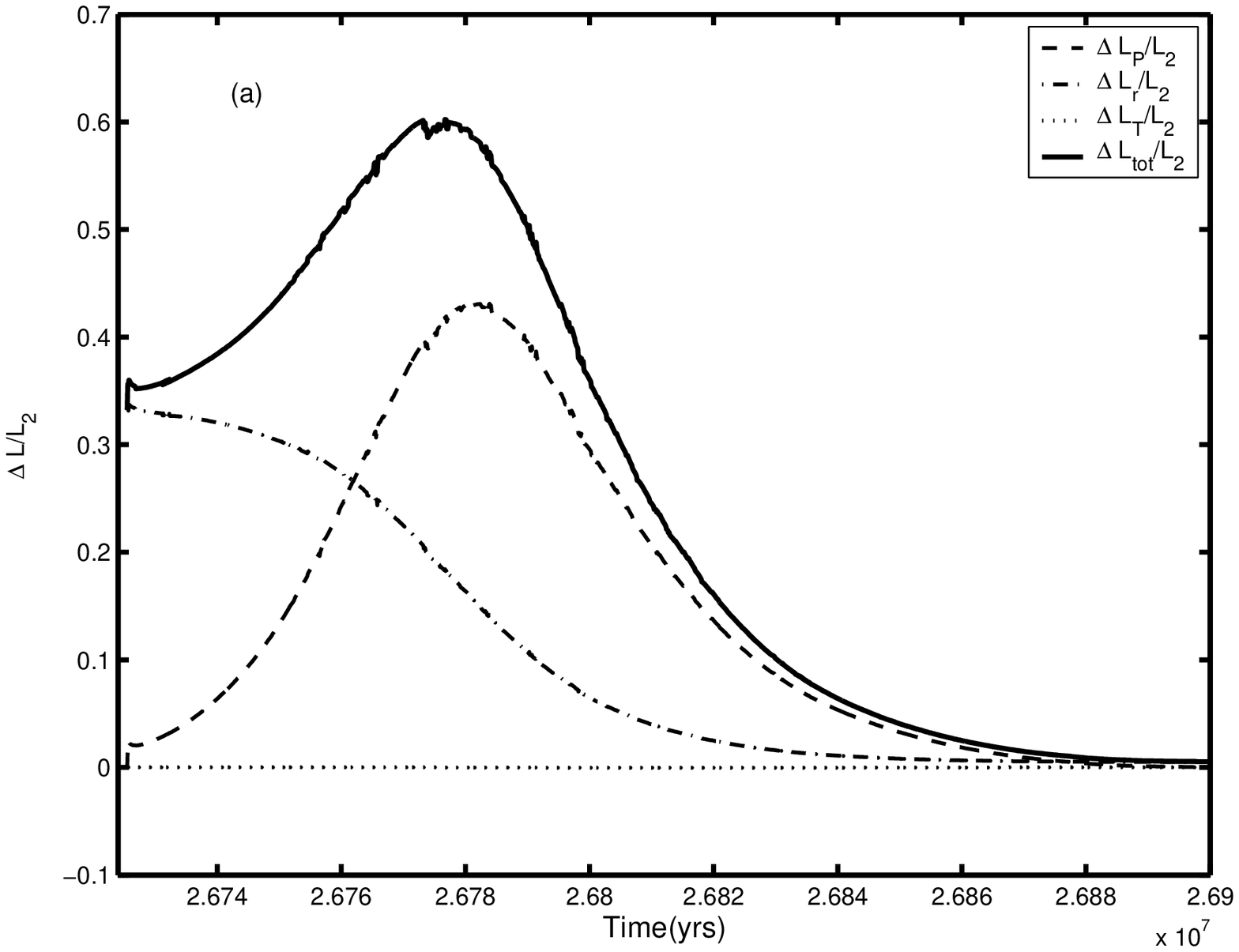}
\includegraphics[width=7.5cm,clip]{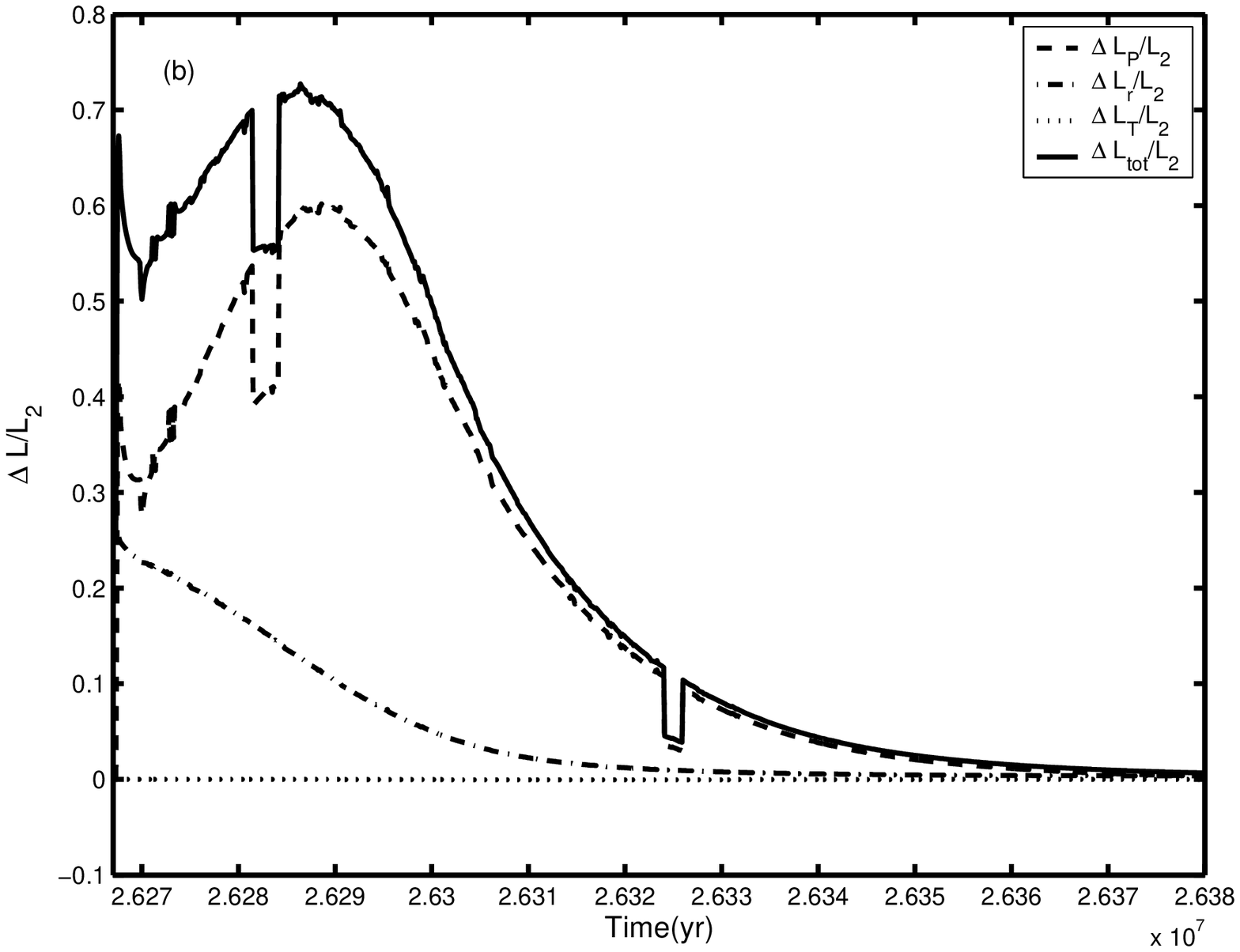}
\caption{Time variation of relative accretion luminosity at
semi-detached stage. Panel (a) represents case 1 and panel (b)
represents case 2. The solid, dotted, dashed and dotted-dashed
curves correspond to the relative accretion luminosity with respect
to total, thermal, potential and irradiative energies,
respectively.} \label{appfig}
\end{figure*}
\begin{figure*}
\centering
\includegraphics[width=7.5cm,clip]{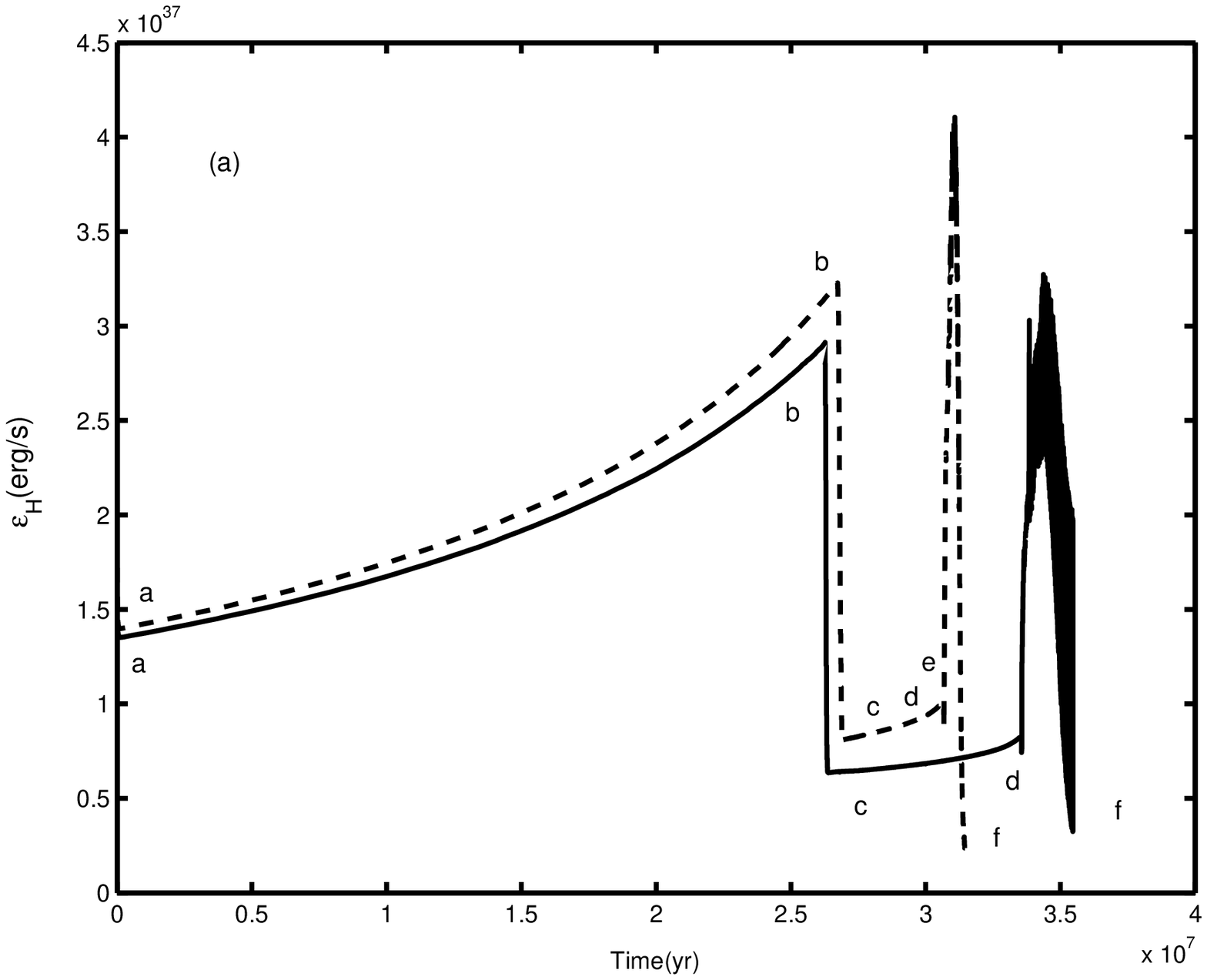}
\includegraphics[width=7.5cm,clip]{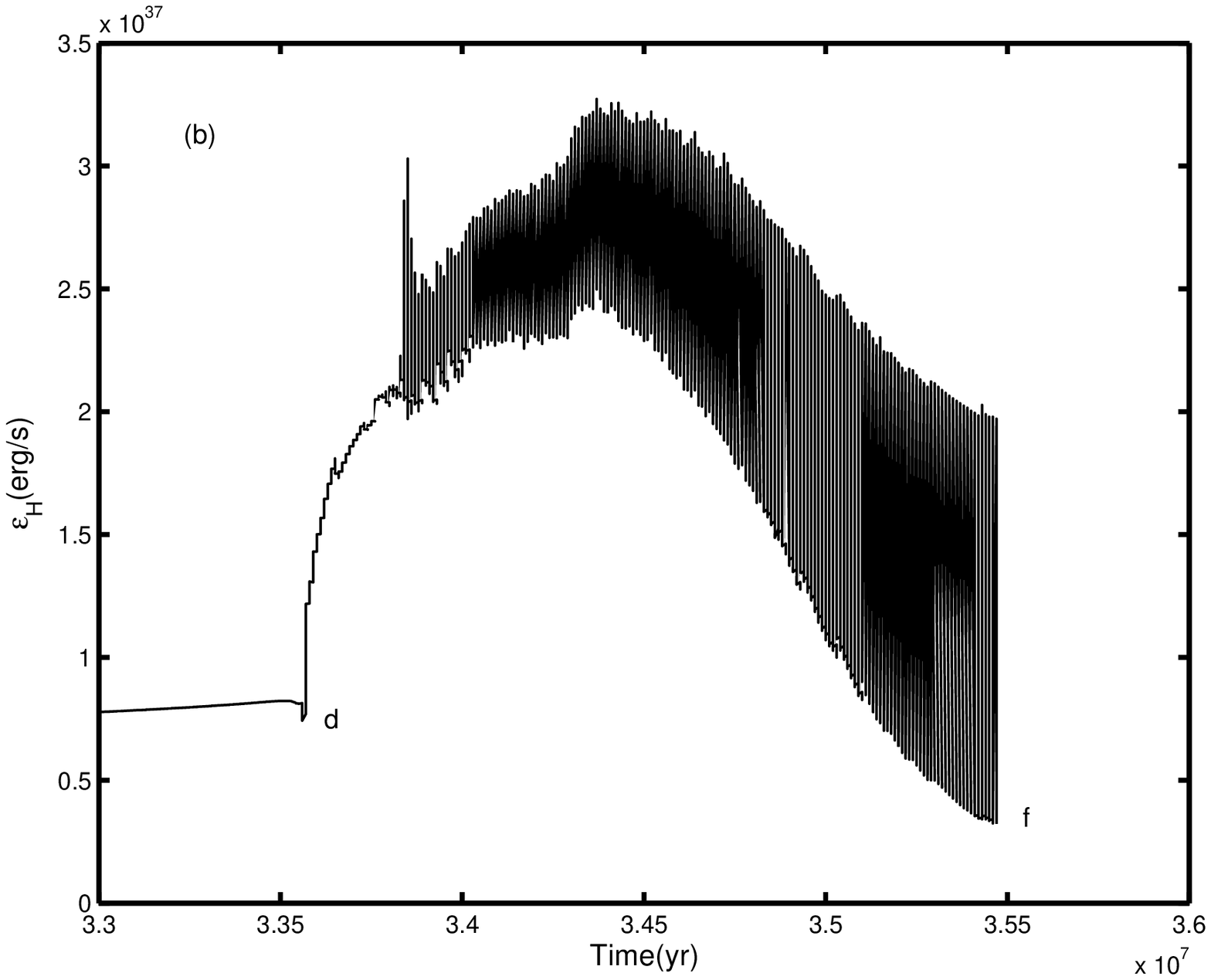}
\caption{ Panel (a): variation in total H-burning generation energy
rate in two cases. Panel (b): time variation in total H-burning
generation energy rate in case 2 after main sequence. The solid
curve represents case 2 and the dashed curve represents case 1. }
\label{appfig}
\end{figure*}
\begin{figure*}
\centering
\includegraphics[width=7.5cm,clip]{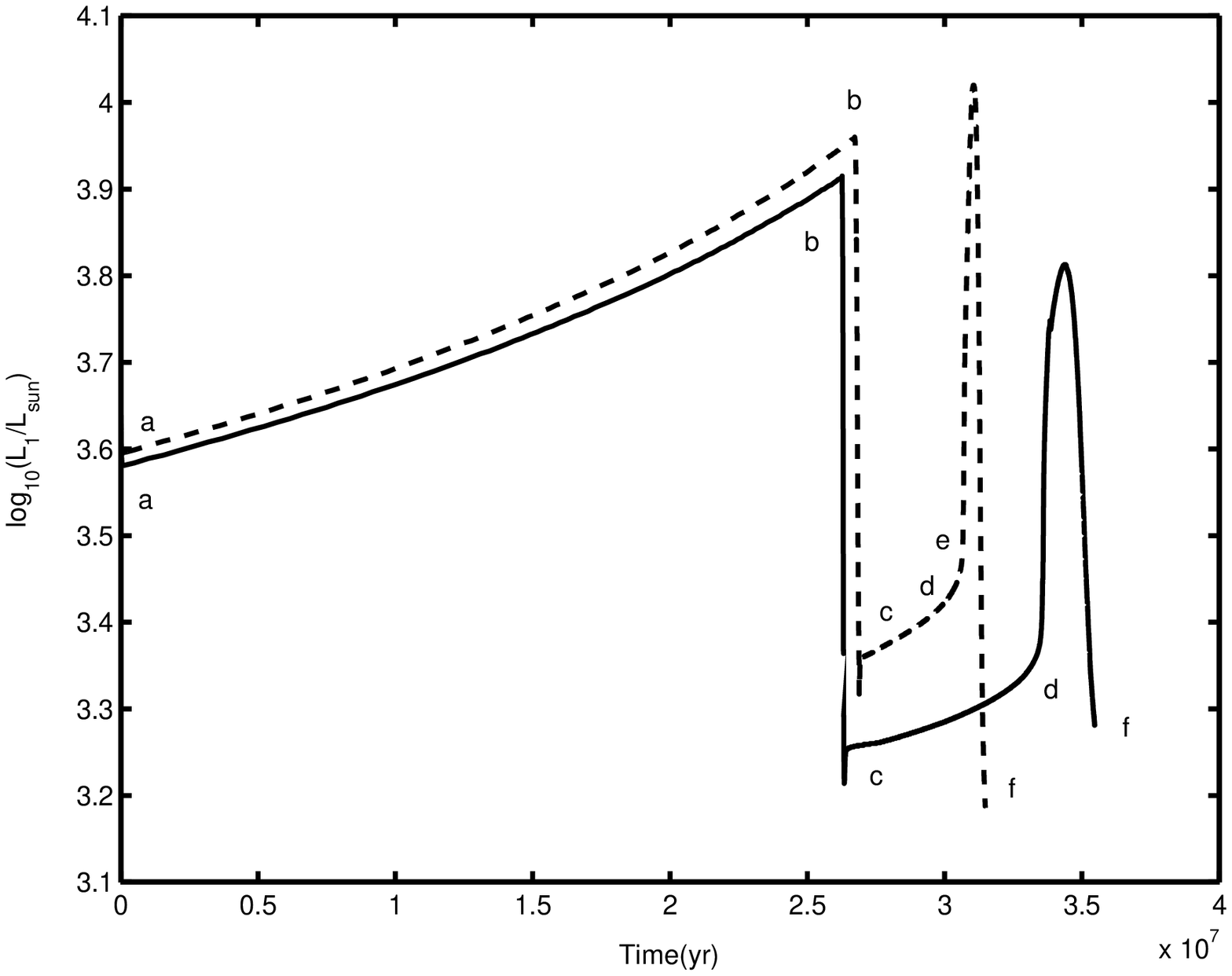}
\includegraphics[width=7.5cm,clip]{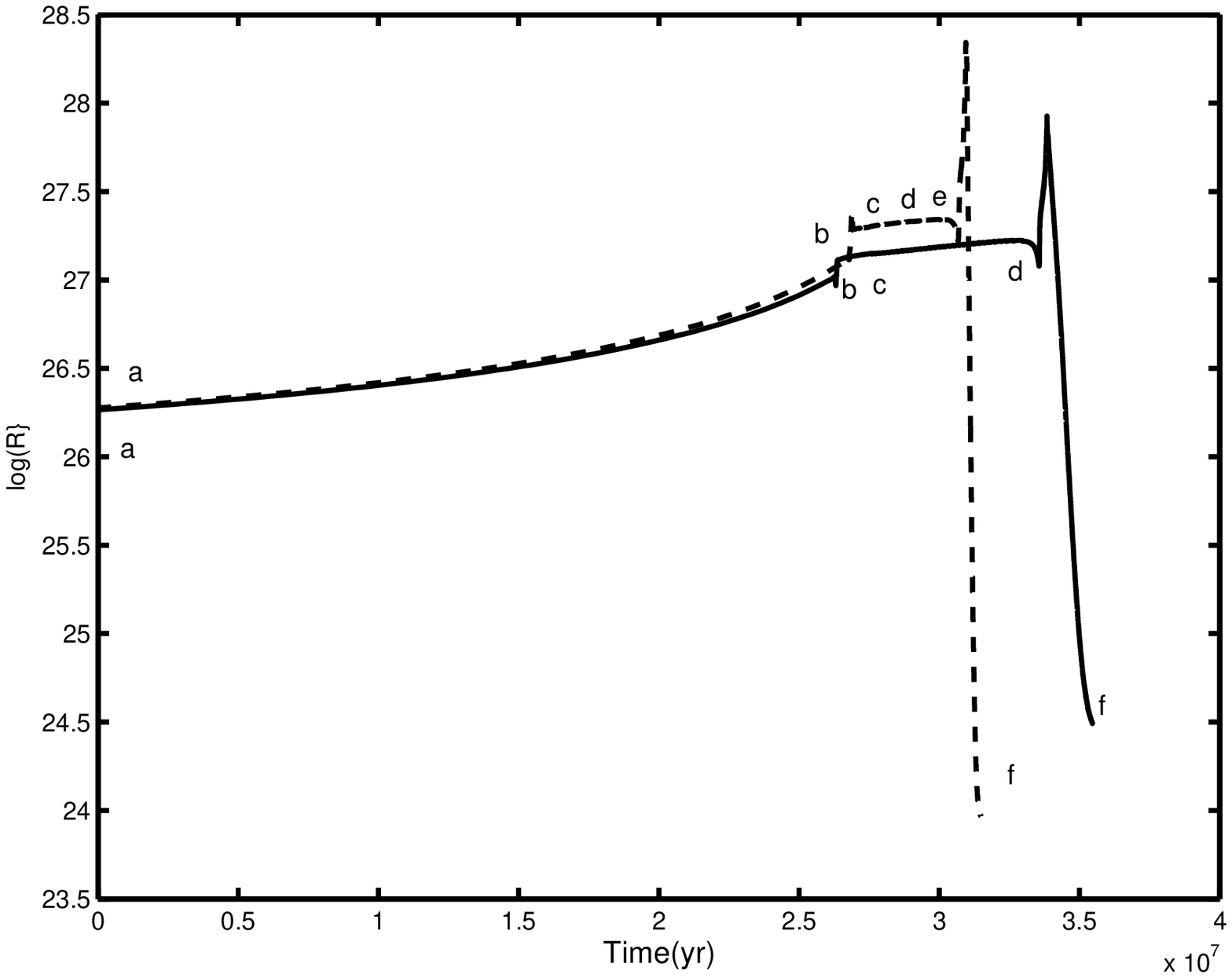}
\caption{ Time-dependent variation in luminosity and equivalent
radius of primary in two cases. The solid and dotted curves have the
same meaning as in Fig. 4. } \label{appfig}
\end{figure*}
\begin{figure*}
\centering
\includegraphics[width=7.5cm,clip]{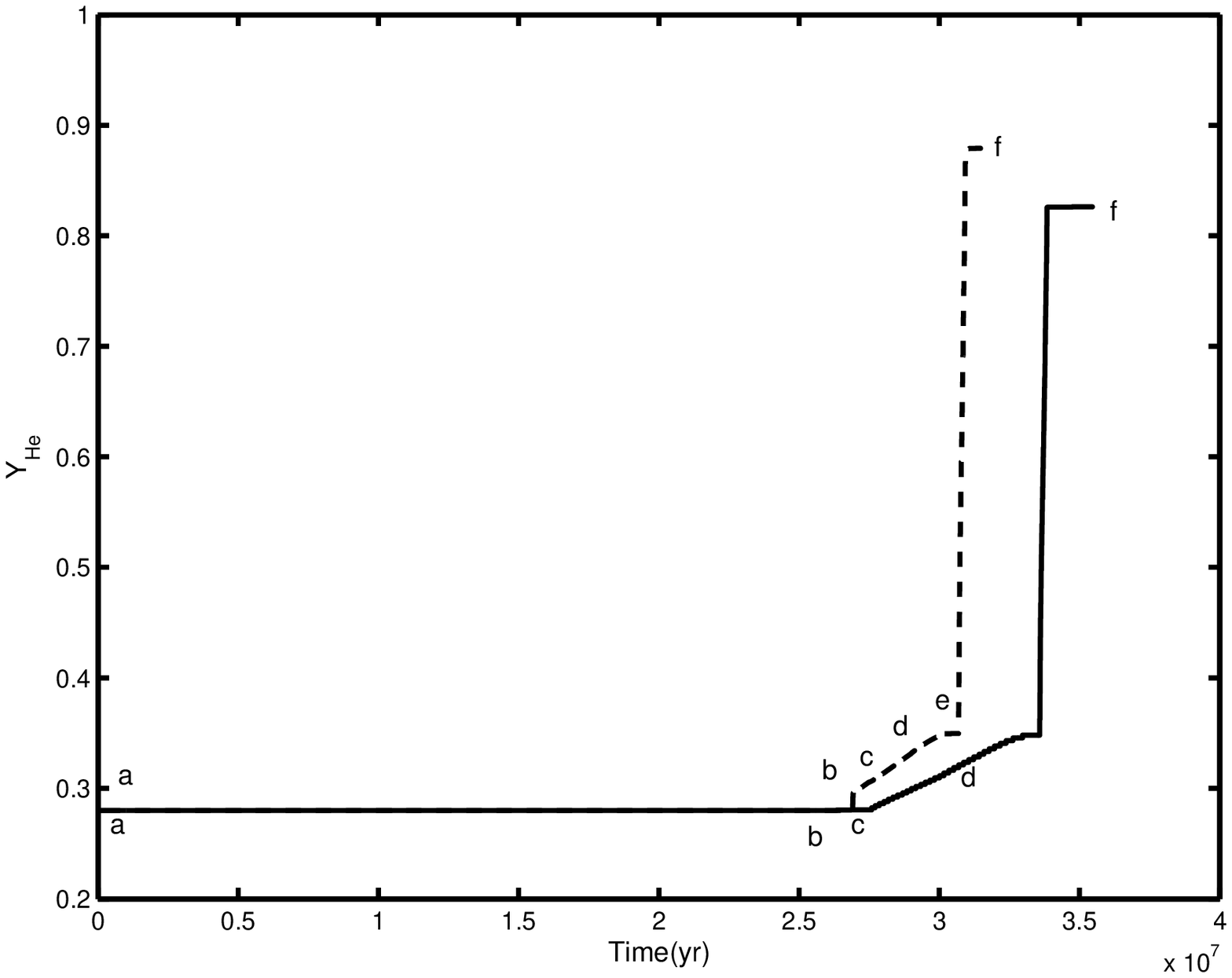}
\caption{Time-dependent variation in surface helium in two cases.
The solid and dotted curves have the same meaning as in Fig. 4.}
\label{appfig}
\end{figure*}

The evolution of the binary system proceeded as follows (cf. Table
1). Evolutionary time, orbital period, mass of two stars,
luminosities and effective temperature of two stars, central and
surface helium mass fraction of the primary, and mean equatorial
rotational velocities of two stars are listed in Table 1. Points a,
b, c, d, e, and f denote the zero-age main sequence, the beginning
of the mass transfer stage, the beginning of H-shell burning, the
end of central hydrogen-burning, the beginning of the central
helium-burning stage, and the end of calculation, respectively. At
the beginning of mass exchange, the luminosity and effective
temperature of the primary component decrease rapidly. The secondary
accretes $6.174M_{\odot}$ for case 1 and $5.502M_{\odot}$ for case 2
during the mass transfer in case A. Because of this mass gain, the
luminosity and the temperature of the secondary go up. When the mass
is transferred from the more massive star to the less massive one,
the separation between the centres of the two components as well as
the orbital period of the system decrease. Some orbital angular
momentum is transformed into the spin angular momentum of both
components, and this process is crucial to model the spin-up of the
accretion star. With mass overflow, the mass of the primary will be
less than that of the secondary. When the mass is transferred from
the less massive star to the more massive one, the separation
between the centres of the two components as well as the orbital
period of the system increases. Some spin angular momenta in both of
the components are transformed into orbital angular momentum. This
physical process results in a longer epilogue after mass transfer.

The equilibrium configuration deviates from spherical symmetry
because of the centrifugal forces and tidal forces. And the deviated
region mainly lies in the outer layer of a star. In fact, the
distorted stellar surface forms the shape of a triaxial ellipsoid. A
distorted isobar surface can be expressed as
\begin{equation}
r=r_{p}[1+f(r)P_{2}(\cos\theta)+g(r)P_{2}^{2}(\cos\theta)cos2\varphi],
\end{equation}
which corresponds to the form of the disturbing potential (Zahn
1992). The coefficients $f(r)$ and $g(r)$  can be defined as
$f(r)=-\frac{C_1\Omega^{2}}{\pi G\varrho}-\frac{C_2
M_{2}}{\pi\varrho D^{3}}$ and $g(r)=\frac{C_2 M_{2}}{2\pi\varrho
D^{3}}$. The quantity $\varrho$ is the mean density of a star with
the mass of $M_{1}$. It was noticed that at the central
hydrogen-burning phase, two parameters  $C_1$ and $C_2$ in Eq. (39)
gain the values of $0.703\pm 0.125$ and $0.491\pm 0.102$,
respectively. This formula indicates that the shapes of the two
components vary with the potentials of the centrifugal force and the
tidal force. The radial deformation is inversely proportional to the
mean density of the component. In order to describe the distortion,
the distribution of the surface rotating velocities of the primary
is illustrated in Fig. 1.  The four panels (a), (b), (c), and (d)
correspond to the evolutive time of 0, $2.3386\times 10^{7}$,
$2.6194\times 10^{7}$, $2.6287\times 10^{7}$ years and corresponding
periods of  $2.776$, $2.760$, $2.746$, and $2.628$ days,
respectively. The rotational velocity rates for the peaks of the
semi-major axes $b$ and $a$ are
$\frac{b}{a}=\frac{v_{b}}{v_{a}}=0.9867, 0.9401, 0.8814,$ and
$0.8664$ in four panels, respectively. The results show that the
surface deformation is intensified with the evolution and
volume-expansion of the primary. The distortion throughout the outer
region of the primary is considerable. The detailed theoretical
models that focus on investigation of the outer regions have
somewhat deviated from the Roche model. The high-order perturbed
potential is required for studying the structure and evolution of
short-period binary systems. Matthews \& Mathieu (1992) examined 62
spectroscopic binaries with A-type primaries and orbital periods
less than 100 days. They concluded that all systems with orbital
periods less than or equal to three days have circular orbits or
nearly circular orbits. Zahn (1977) and Rieutord \& Zahn (1997) have
shown how binary synchronization and circularization result from
tidal dissipation. Based on smoothed particle hydrodynamics (SPH)
simulation, Renvoiz$\acute{e}$ et al.(2002) have quantified the
geometrical distortion effect due to the tidal and rotational forces
acted on the polytropic secondaries of semi-detached binaries. They
suggest that the tidal and rotational distortion on the secondary
may not be negligible, for it may reach observable levels of $\sim
10\%$ on the radius in specific cases of polytropic index and mass
ratio. Georgy et al.(2008) display that various effects of the
rotation on the surface of a 20$M_{\odot}$ star at a metallicity of
$10^{-5}$ and at $\sim 95\%$ of the critical rotation velocity. They
point out that the star becomes oblate with an equatorial-to-polar
radius ratio $\frac{R_{eq}}{R_{pol}}\simeq 1.3$. These results agree
closely with ours.

The variation relative gravitational accelerations, the tidal force,
and the ratio of $f_{cen}/f_{tid}$ on the surface of the primary
under the coordinate $\theta$ and $\varphi$ at the beginning of mass
overflow are shown in Fig. 2.  The quantities $g_{r}$, $g_{\theta}$,
and $g_{\varphi}$ are the three components of gravitational
acceleration. The six panels (a), (b), (c), (d), (e), and (f)
represent the distribution of $g_{r}/g$, $g_{\theta}/g$,
$g_{\varphi}/g$, $g_{tot}/g$, $f_{tid}$, and $f_{cen}/f_{tid}$,
respectively.  The quantity $g$ equals the gravitational
acceleration of the corresponding equivalent sphere
($g=\frac{GM_{1}}{r_{p}^{2}}$). When the joint effect of rotation
and tide is considered, the gravitational accelerations are
different from those in the conventional model. Gravitational
acceleration generally has three components.

It is shown in panel (a) that the relative quantity
$\frac{g_{r}}{g}$ reaches the maximum value of $1.048$ at the two
polar points and drops to the minimum value $0.6987$ on the
equatorial plane because the inward tidal force acts on the primary
and causes the polar radius to become shorter. The tidal and
centrifugal forces pull the primary outwards and change
gravitational accelerations greatly on the equatorial plane.
Furthermore, the maximum value is $0.9486$ and the minimum value is
$0.6987$ on the equatorial plane. The lower values are at the peak
of the longest axis $a$ and the higher values are at the peak of the
axis $b$. The relative quantity $\frac{g_{\theta}}{g}$
 reaches the maximum value of $0.20399$ at the point of
$\theta=\frac{k\pi}{2}+\frac{\pi}{4}; \varphi=k \pi$, $k=0,1$ and
vanishes at the two polar points and on the equatorial plane in
panel (b). It can be seen that a secondary maximal value of
$0.10214$ exists at point of $\theta=\frac{k\pi}{2}+\frac{\pi}{4};
\varphi=k \pi+\frac{\pi}{2}$, $k=0,1$.  The relative quantity
$\frac{g_{\varphi}}{g}$ reaches the maximum value of  $0.1050$ at
point $\theta=\frac{\pi}{2}; \varphi=\frac{k\pi}{2}+\frac{\pi}{4}$,
k=0,1,2,3 and decreases to zero at the point of
$\varphi=\frac{k\pi}{2}$, $k=0,1,2,3$ in panel (c). The total
gravitational acceleration at the surface of the primary is shown in
panel (d). Its distribution is similar to the one of
$\frac{g_{r}}{g}$ because the radial component is the maximum value.
It is noticed that, as expected, the average gravitational
acceleration of the rotating model is less than for the non-rotating
model. It can be observed that the tidal force reaches the highest
value of $340.05 cm/s^{2}$ at the point of $\theta=\frac{\pi}{2};
\varphi=k\pi$, $k=0,1$  and decreases to the lowest of $136.24
cm/s^{2}$ at the two polar points. The quantity $f_{cen}/f_{tid}$
reaches the maximum value of $2.4905$ at the point of
$\theta=\frac{\pi}{2}; \varphi=k\pi+\frac{\pi}{2}$, $k=0,1$ and
reaches the secondary maximal value of  $1.2445$ at the point of
$\theta=\frac{\pi}{2}; \varphi= k\pi$, $k=0,1$. The results show
that the effect produced by tidal distortion is lower in comparison
with what is produced by rotational distortion on the equatorial
plane. However, with the mass conversion, the opposite situation can
emerge. It is concluded that tidal distortions are related to the
mass ratio of the secondary to the primary. These results suggest
that rotation and tide have strong influences on the stellar
surface. They modify the gravity and change the
spherically-symmetric shape into the triaxial ellipsoid shape.
Furthermore, the stellar structure equations are basically revised
due to the distribution of the relative quantity in the outer
region.

According to the Von Zeipel theorem, the mass loss due to stellar
winds should be proportional to local effective gravity. Polar
ejection is intensified by the tidal effect. The higher gravity at
the peak of the axis $b$ makes it hotter. The ejection of an
equatorial ring may be favoured by both the opacity effect and the
higher temperature at the peak of the semi-axis $b$. This effect is
called the $g_{e}(\theta,\varphi)$-effect in this paper. It is
predicted that the $g_{e}(\theta,\varphi)$-effect is as important as
the $g_{e}$-effect suggested by Maeder (1999) and Maeder \&
Desjacques (2001). The shapes of planetary nebulae that deviate from
spherical symmetry (axisymmetrical one in particular) are often
ascribed to rotation or tidal interaction (Soker 1997). Frankowski
and Tylenda (2001) suggest that a mass-losing star can be noticeably
distorted by tidal forces, thus the wind will exhibit an intrinsic
directivity and may be globally intensified. Interestingly enough,
the group of the B[e] stars shows a two-component stellar wind with
a hot, highly ionized, fast wind at the poles and a slow, dense,
disk-like wind at the equator (Zickgraf 1999). Maeder and Desjacques
(2001) have noticed that the polar lobes and skirt in $\eta$ Carinae
and other LBV stars may naturally result from the $g_{eff}$ and
$\kappa$-effects. Langer et al. (1999) have shown that giant LBV
outbursts depend on the initial rotation rate. Tout and Eggleton
(1988) proposed a formula according to which the tidal torque would
enhance the mass-loss rate by a factor of
$1+B\times(\frac{R}{R_{RL}})^{6}$, where $B$  is a parameter free to
be adjusted (ranging from $5\times10^{2}$ to $10^{4}$).  Mass loss
and associated loss of angular momentum are anisotropic in rotating
binary stars. The theories for describing the mass loss and angular
momentum loss from stellar winds should be altered partly in future
work.

The time variation of relative accretion luminosity at the
semi-detached stage is shown in Fig. 3. The two panels (a) and (b)
correspond to cases 1 and 2, respectively. The figure shows that the
release of transferred thermal energy approaches zero, which
indicates that the transferred thermal energy can be ignored in the
two cases. The irradiation energy plays an important role in the
early stage of mass overflow and attenuates at the subsequent stage,
which can be explained by the luminosity of the primary decreasing
rapidly and the luminosity of the secondary increasing with mass
transfer gradually. The transferred potential can exceed the
irradiation energy as the mass transfer rate grows. The total
accretion luminosity in case 2 is higher than the one in case 1
because the potentials in the two cases are different. From panel
(b), it can be seen that the curve of the accretion luminosity
fluctuates, indicating that the mass transfer process is unstable.

The total H-burning energy-generation rates of the primary in the
two cases are shown in Fig. 4. Panel (b) shows the H-burning
energy-generation rate in case 2 after the main sequence. From the
difference between curves in panel (a), it is noticed that the
effect of rotation causes the total H-burning energy-generation rate
lower. As a result, the evolutive time in the main-sequence stage
gets longer (cf. Table 1). Moreover, the larger fuel supply and
lower initial luminosity of the rotating stars help to prolong the
time which they spend on the main sequence (Heger \& Langer 2000b).
The lifetime extension in rotating binary star at the main-sequence
stage can also be illustrated according to Suchkov (2001). Their
results show that the age-velocity relation (AVR) between F stars in
the binary system is different from the one between ``truly single''
F stars. The discrepancy between the two AVRs indicates that the
putative binaries are, on average, older than similar normal single
F stars at the same effective temperature and luminosity. It is
speculated that this peculiarity comes from the impact of the
interaction of components in a tight pair on stellar evolution,
which results in the prolonged main-sequence lifetime of the primary
F star.  Moreover, no central helium-burning stage exists for case 2
(cf. Table 1). From panel (b), it can be seen that the
energy-generation rate of the primary vibrates at the H-shell
burning stage in case 2. These facts suggest that the burning of
H-shell is unstable in case 2. The reason lies in the centrifugal
force reducing the effective gravity at the stellar envelope. The
luminosity and surface temperature there decrease (Kippenhahn 1977;
Langer 1998; Meynet and Maeder 1997). Thus, the shell source becomes
cooler, thinner, and more degenerated as the He core mass increases.
As the hydrogen shell becomes instable, the thickness
$\frac{D}{r_{s}}$ and surface temperature are $\sim0.203$ and
$1.1885\times10^{4}K$, respectively. This physical condition leads
to thermal instability (Yoon et al., 2004), and the H-shell source
experiences slight oscillation. It is well known that the
energy-generation rate is proportional to temperature and density
($\varepsilon \propto \rho T^{n}$);  therefore, the curve of the
H-shell energy-generation rate fluctuates.

The time-dependent variation in the luminosity and the equivalent
radius of the primary in the two cases are illustrated in Fig. 5.
Because the rotating star has a lower energy-generation rate, the
luminosity of the primary is lower, which is the consequence of
decreased central temperature in rotating models due to decreased
effective gravity (Meynet and Maeder 1997). Then, the primary
expands slowly in case 2. It is observed that case 1 reaches point b
at $t=2.6725\times10^{7}yr$, while case 2 reaches point b at
$t=2.6267\times10^{7}yr$. The initiation time of mass transfer for
case 2 is advanced by about $\sim 1.71\%$. Similarly, numerical
calculation by Petrovic et al.(2005b) shows the radius of the
rotating primary increases faster than that of the non-rotating
primary due to the influence of centrifugal forces. Their results
also show that mass transfer of Case A starts earlier in rotating
binary system, which is consistent with ours. If the rotating star
is still treated as a spherical star, the initiation time of mass
overflow should be later than that in the non-rotational case.
Actually, because of the distortion by rotation and tide, the time
for mass overflow may be extended.  Therefore, it is very important
to investigate distortion in close binary systems.

The time-dependent variation in the helium compositions at the
surface of the primary is illustrated in Fig. 6. The H-shell burning
begins at $t=2.6854\times10^{7}yr$ in case 1 while at
$t=2.6329\times10^{7}yr$ in case 2 (cf. Table 1). Therefore, the
initiation time of H-shell burning is advanced by
$1.71\times10^{5}yr$. Moreover, the helium composition at the
surface of the primary is $0.280051$ at point c, suggesting that the
diffusion process progresses slowly in a rotating star. Cantiello et
al.(2007) also indicate that rotationally induced mixing before the
onset of mass transfer is negligible, in contrast to typical $O$
stars evolving separately; hence, the alteration of surface
compositions depends on both initial mass and rotation rates. The
sample of the OB-type binaries with orbital periods ranging from one
to five days by Hilditch et al. (2005) shows enhanced N abundance up
to $0.4$ dex. Langer et al. (2008) have discovered that for the same
binary system, but with the initial period of six days instead of
three days, its mass gainer is accelerated to a rotational velocity
of nearly $500km s^{-1}$, which produces an extra nitrogen
enrichment from more than a factor two to about 1 dex in total.
Because there is no central helium-burning phase for case 2, the
diffusion process can be neglected in the interior region of the
primary after the main sequences.

\section{Conclusions}

The main achievements of this study may be summarised as follows.

(a) The distortion throughout the outer layer of the primary is
considerable. The detailed theoretical models that investigate the
outer regions of the two components have deviated somewhat from the
lowest approximation of the Roche model. The high-order perturbing
potential is required especially in the investigation of the
evolution of short-period binary system.

(b) The equilibrium structures of distorted stars are actually
triaxial ellipsoids. A formula describing rotationally and tidally
distorted stars is presented. The shape of the ellipsoid is related
to the mean density of the component and the potentials of
centrifugal and tidal force.

(c)The radial components of the centrifugal force and the tidal
force cause the variation in gravitation. The tangent components of
the centrifugal force and the tidal force cannot be equalized and,
instead, they change the shapes of the components from perfect
spheres to triaxial ellipsoids. Mass loss and associated angular
momentum loss are anisotropic in rotating binary stars. Ejection is
intensified by tidal effect. The ejection of an equatorial ring may
be favoured by both the opacity effect and the higher temperature at
the peak of semi-axis $b$. This effect is called the
$g_{e}(\theta,\varphi)$-effect in this paper.

(d) The rotating star has an unstable H-burning shell after the main
sequence. The components expand slowly due to their lower
luminosity. If the components are still treated as spherical stars,
some important physical processes can be ignored.

\begin{acknowledgements}
We are grateful to Professor Norbert Langer and Dr.
St$\acute{e}$phane Mathis for their valuable suggestions and
insightful remarks, which have improved this paper greatly. Also we
thank Professor Norbert Langer for his kind help in improving our
English.
\end{acknowledgements}

\end{document}